\documentclass[prl,twocolumn,superscriptaddress,citeautoscript,showpacs,amsart,longbibliography]{revtex4-1}

\usepackage{graphicx}
\usepackage{multirow}
\usepackage{color}
\usepackage{bm}
\usepackage{times}
\usepackage{amsmath,bm,amsfonts}
\usepackage{dcolumn}
\usepackage{graphicx}
\usepackage{latexsym}
\usepackage{gensymb}
\usepackage{verbatim}

\usepackage{ulem} % to strike things out \normalem % usual emph
\usepackage{braket}

\usepackage{cancel}

\def\ket#1{|#1\rangle }
\def\bra#1{\langle #1 |}

\begin{document}
\title{
Macroscopically degenerate localized zero-energy states of quasicrystalline bilayer systems in strong coupling limit
}

\author{Hyunsoo \surname{Ha}}

\affiliation{Department of Physics and Astronomy, Seoul National University, Seoul 08826, Korea}

%\affiliation{Imae Elementary school, Bundang-gu, Seongnam-si, Gyeonggi-do 13567, Korea}

\author{Bohm-Jung \surname{Yang}}
\email{bjyang@snu.ac.kr}
\affiliation{Department of Physics and Astronomy, Seoul National University, Seoul 08826, Korea}

\affiliation{Center for Correlated Electron Systems, Institute for Basic Science (IBS), Seoul 08826, Korea}

\affiliation{Center for Theoretical Physics (CTP), Seoul National University, Seoul 08826, Korea}

\date{\today}

\begin{abstract}
When two identical two-dimensional (2D) periodic lattices are stacked in parallel after rotating one layer by a certain angle relative to the other layer, the resulting bilayer system can lose lattice periodicity completely and become a 2D quasicrystal.
Twisted bilayer graphene with $30\degree$ rotation is a representative example.
We show that such quasicrystalline bilayer systems generally develop macroscopically degenerate localized zero-energy states (ZESs) in strong coupling limit 
where the interlayer couplings are overwhelmingly larger than the intralayer couplings.
The emergent chiral symmetry in strong coupling limit and aperiodicity of bilayer quasicrystals guarantee the existence of the ZESs.
The macroscopically degenerate ZESs are analogous to the flat bands of periodic systems, in that both are composed of localized eigenstates, which give divergent density of states.
For monolayers, we consider the triangular, square, and honeycomb lattices, comprised of homogenous tiling of three possible planar regular polygons: the equilateral triangle, square, and regular hexagon.
We construct a compact theoretical framework, which we call the quasiband model, that describes the low energy properties of bilayer quasicrystals and counts the number of ZESs using a subset of Bloch states of monolayers. 
We also propose a simple geometric scheme in real space which can show the spatial localization of ZESs and count their number.
Our work clearly demonstrates that bilayer quasicrystals in strong coupling limit are an ideal playground to study the intriguing interplay of flat band physics and the aperiodicity of quasicrystals.
\end{abstract}

\maketitle

%\tableofcontents

\textit{Introduction.---} 
Flat bands with divergent density of states are ideal playgrounds to explore strong correlation physics~\cite{flat_highTsc,flat_FQHE1,flat_FQHE2,flat_FQHE3,flat_FQHE4,flat_FQHE5, flat_Wigner}.
Recent discovery of twisted bilayer graphene (TBG) has provided a new platform for studying flat bands. 
In TBG, where two graphene layers are stacked with relative inplane rotation by an angle $\theta$, 
the electronic properties change dramatically depending on $\theta$. 
Although TBG is generally not periodic due to incommensurate interlayer potentials, when $\theta$ is small ($\theta<10\degree$), 
its low energy properties can be well described by Moire effective theory with translational invariance. 
Especially, when $\theta$ takes special values, called the magic angles, the Dirac cones of graphene layers coupled by the Moire superlattice potential are strongly renormalized and turn into nearly flat bands ~\cite{magicangle, CastroNeto_TBGelectronicstructure, CastroNeto_TBGcontinuum,Kindermann_continuum,koshino_moon_optical}, 
which induce intriguing strong correlation physics~\cite{tbg_PJHmott,tbg_PJHsc,tbg_sc}.

Another interesting limit of TBG is when $\theta=30\degree$ exactly. 
In this case, the system  completely loses translational symmetry~\cite{no_translational_sym1,no_translational_sym2}, but instead, develops a quasicrystalline order with 12-fold rotation symmetry.
Such an extrinsic quasicrystal of TBG, composed of two periodic graphene layers~\cite{stampfli,stampfli2}, was recently realized in experiment~\cite{ahn}.
Angle-resolved photoemission spectroscopy has shown many replicas of Dirac cones with 12-fold rotational symmetry in TBG quasicrystal, arising from the interlayer coupling with the quasicrystalline order~\cite{ahn}. 
However, most of the previous studies are limited to the weak coupling limit where the aperiodic interlayer hopping is much smaller than the intralayer one, 
so that the low energy properties of the TBG quasicrystal are governed by massless Dirac fermions as in monolayer graphene~\cite{Son, Sungbin}.

%%%%%%%%%%%%%%%%%%%%%%%%%%%%%%%%%%%%%%%%   FIGURE   %%%%%%%%%%%%%%%%%%%%%%%%%%%%%%%%%%%%%%%%%%%%%%%%%%%%%%
\begin{figure}
\includegraphics[width=0.5\textwidth]{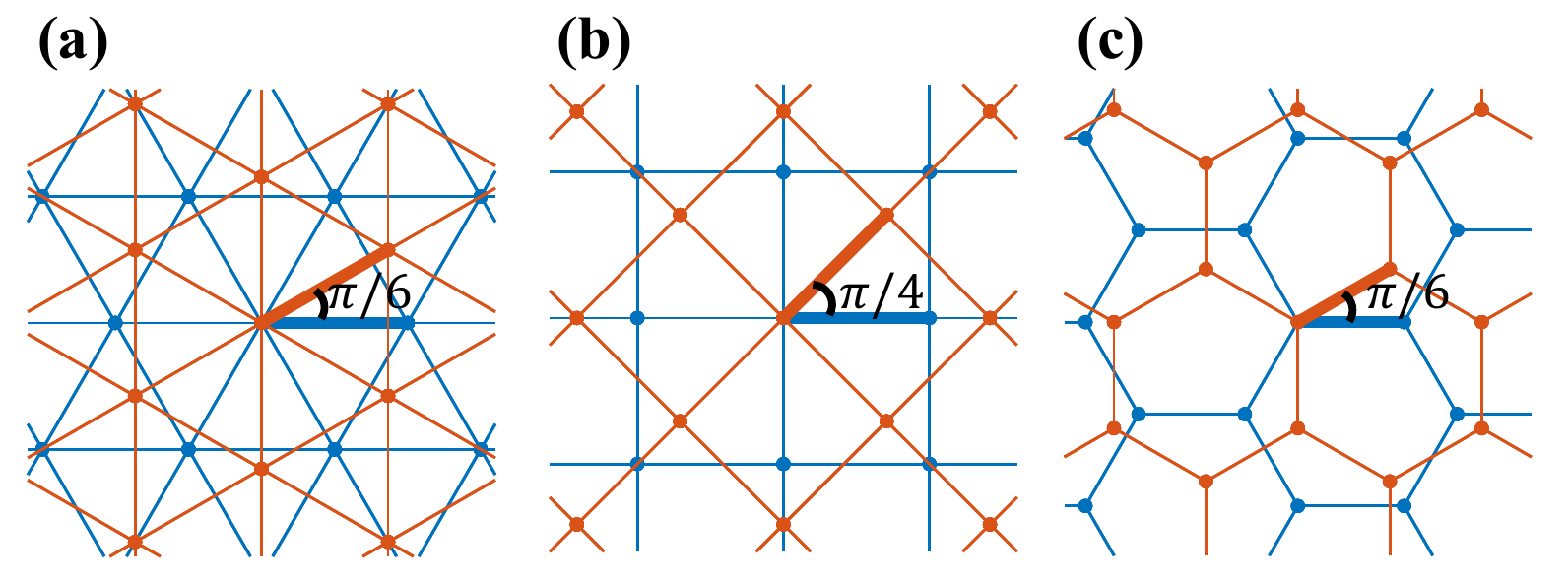}
\caption{ Lattice structures of
(a) twisted bilayer triangular lattice (TBTL) quasicrystal with $\theta=30\degree$, 
(b) twisted bilayer square lattice (TBSL) quasicrystal with $\theta=45\degree$,
(c) twisted bilayer honeycomb lattice quasicrystal with $\theta=30\degree$.
}
\label{fig:structures}
\end{figure}
%%%%%%%%%%%%%%%%%%%%%%%%%%%%%%%%%%%%%%%%   FIGURE   %%%%%%%%%%%%%%%%%%%%%%%%%%%%%%%%%%%%%%%%%%%%%%%%%%%%%%

In this Letter, we unveil the low energy properties of quasicrystalline bilayer systems
in strong coupling limit where the intralayer hopping is negligibly
small compared to the aperiodic interlayer hopping. 
To establish the general properties of 2D quasicrystalline bilayers beyond the TBG with $\theta=30\degree$, 
we take the triangular, square, and honeycomb lattices as a monolayer structure, and compose the corresponding bilayer quasicrystals with the rotation angle  
$\theta=30\degree$, $45\degree$, and $30\degree$, respectively [see Fig.~\ref{fig:structures}].
We note that these monolayers are made of homogenous tiling of three allowed planar regular polygons, i.e., the equilateral triangles, squares, and regular hexagons.
Surprisingly, we find that quasicrystalline bilayers in strong coupling limit generally host macroscopically degenerate localized zero-energy states (ZESs).
The chiral symmetry related to layer degrees of freedom, which emerges in strong coupling limit, 
guarantees the presence of the ZESs.
We note that the macroscopically degenerate ZESs are similar to the flat bands of periodic lattices, both of which induce divergent density of states (DOS).
Hence the quasicrystalline bilayer in strong coupling limit provides a fascinating platform to study the interplay of the flat energy dispersion and the quasicrystalline order,
which is distinct from the weak coupling physics dominated by masselss Dirac fermions.
%We note that similar macroscopically degenerate ZESs was predicted and experimentally observed~\cite{pseudogap, pseudogap_NMR,pseudogap_Xray} in intrinsic quasicrystals such as the bipartite Penrose lattice ~\cite{penrose, penrose_ratio, penrose_perp, penrose_localize, penrose_E, penrose_E2, penrose_E3, penrose_hubbard}. 
%Our work clearly demonstrates that the physics of the macroscopically degenerate ZESs is not limited to the intrinsic quasi-crystals but generally observable in extrinsic quasicrystals composed of two layers of periodic 2D lattices.

{\it Twisted bilayer triangular lattice (TBTL) quasicrystal.---}
We first consider the quasicrystal comprised of twisted bilayer triangular lattices (TBTLs) with $\theta=30\degree$.
For comparison with TBG quasicrystal, we construct a model where a $p_z$ orbital is placed at each lattice site.
The full Hamiltonian is given by $H=H_1+H_2+H_{12}$ where $H_{i=1,2}$ describes intralayer hoppings in the layer $i$ while $H_{12}$
indicates interlayer hoppings. In strong coupling limit, we neglect $H_{i=1,2}$ whose influence is discussed later.
The resulting Hamiltonian is
\begin{align}
\label{eq.LatticeH}
H=-\sum_{\mathbf{R}_{1},\mathbf{R}_{2}}T(\mathbf{R}_{1}-\mathbf{R}_{2})\ket{\mathbf{R}_{1}}\bra{\mathbf{R}_{2}} +h.c.
\end{align}
where $\ket{\mathbf{R}_{1}}$ ($\ket{\mathbf{R}_{2}}$) indicates the $p_z$ orbital located at the position $\mathbf{R}_{1}$ ($\mathbf{R}_{2}$) in the layer 1 (layer 2).
$T(\mathbf{R}_{1}-\mathbf{R}_{2})$ is the transfer integral between two orbitals at $\mathbf{R}_{1}$ and $\mathbf{R}_{2}$, respectively,
generally taking the following form \cite{slaterkoster},
\begin{align}
\label{eq.transfer integral}
-T(\mathbf{R})&=V_{pp\pi}\left[1-\left(\frac{\mathbf{R}\cdot\mathbf{e}_z}{R}\right)^2\right]+V_{pp\sigma}\left(\frac{\mathbf{R}\cdot\mathbf{e}_z}{R}\right)^2,\nonumber\\
V_{pp\pi}&=V_{pp\pi}^0 e^{(-R-a_0)/r_{\star}}, V_{pp\sigma}=V_{pp\sigma}^0 e^{(-R-d_0)/r_{\star}},
\end{align}
where $a_0$ is the inplane lattice constant, $d_0$ denotes the vertical distance between two layers, $\mathbf{R}$ is the relative displacement vector between two atoms, $R=|\mathbf{R}|$, $\mathbf{e}_z$ denotes the unit vector in the $z$-direction.
$V_{pp\pi}$ ($V_{pp\sigma}$) indicates the transfer integral between nearest-neighbor $p_z$ orbitals forming $\pi$-bonding ($\sigma$-bonding). 
The length scale $r_\star$ determines the range of hopping transfer. 
For comparision to TBG quasicrystal, we assume $a_0\approx0.142\mathrm{nm}$, $d_0\approx0.335\,\mathrm{nm}$, $r_\star=r_0\approx 0.0453\,\mathrm{nm}$, $V_{pp\pi}^0\approx-2.6\,\mathrm{eV}$, and $V_{pp\sigma}^0\approx0.48 \,\mathrm{eV}$. Here $r_0$ indicates the interaction range in graphene~\cite{graphenetransfer,hofstadter,Son}.

Interestingly, the strong coupling Hamiltonian in Eq.~(\ref{eq.LatticeH}) has chiral symmetry related to layer degrees
represented by the operator $U=\sum_{\mathbf{R_{1}}}\ket{R_{1}}\bra{R_{1}}-\sum_{\mathbf{R_{2}}}\ket{R_{2}}\bra{R_{2}}$. 
As $\{H,U\}=0$, for an eigenstate $\ket{\psi}$ satisfying $H\ket{\psi}=E\ket{\psi}$, 
one can always find another eigenstate $U\ket{\psi}$ with the energy $-E$ so that the energy spectrum is particle-hole symmetric.

%%%%%%%%%%%%%%%%%%%%%%%%%%%%%%%%%%%%%%%%   FIGURE   %%%%%%%%%%%%%%%%%%%%%%%%%%%%%%%%%%%%%%%%%%%%%%%%%%%%%%
\begin{figure}
\includegraphics[width=0.5\textwidth]{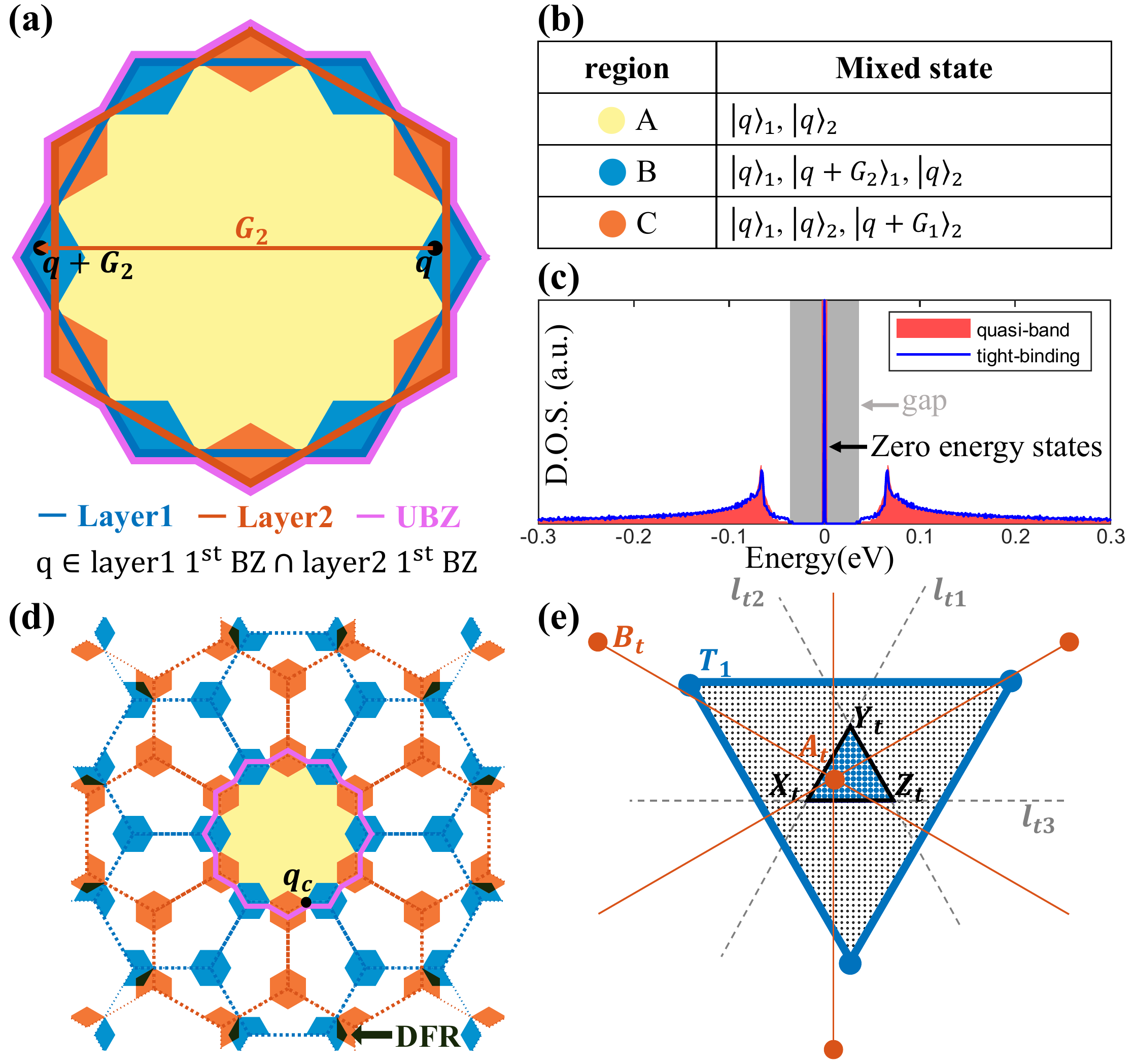}
\caption{
(a) The first Brillouin zone (BZ) of the layer 1 and 2, and the union of BZs (UBZ) taken as the effective BZ (EBZ).
EBZ is divided into three regions $A$ (yellow), $B$ (blue), $C$ (orange).
(b) Bloch states coupled in each region.
(c) Density of states (DOS) from the full tight-binding model and the quasiband model.
(d) Regions $B$ and $C$ illustrated in the extended BZ scheme. The dark filled region (DFR) where regions $B$ and $C$ overlap can couple ZESs in the EBZ.
(e) Real space geometric consideration for ZESs. Orange dots denote the layer 2 where ZESs are confined. 
The small triangle with blue dots inside indicates the possible center positions of localized ZESs with maximum amplitude. 
}
\label{fig:triangle}
\end{figure}
%%%%%%%%%%%%%%%%%%%%%%%%%%%%%%%%%%%%%%%%   FIGURE   %%%%%%%%%%%%%%%%%%%%%%%%%%%%%%%%%%%%%%%%%%%%%%%%%%%%%%

{\it Quasiband model.---}
Though translational invariance is lost in the bilayer quasicrystal, its low energy properties can be
successfully described by the Bloch states of monolayers 
$\ket{\mathbf{k}_j}_j=\frac{1}{\sqrt{N}}\sum_{\mathbf{R}_{j}} e^{i\mathbf{k}_j\cdot\mathbf{R}_{j}}\ket{\bf{R_{j}}}$ $(j=1,~2)$
where $\mathbf{k}_j$ indicates the momentum in the Brillouin zone (BZ) of the $j$th layer and N is the number of lattice sites per layer. 
The subscript $j$ under ket (or bra) is the layer index of the Bloch basis.
The matrix element of the Hamiltonian under the Bloch basis is \cite{meleTBG, magicangle, koshinointerlayer}
\begin{multline}
\label{eq.generalumklapp}
{}_2\bra{\mathbf{k}_2}H\ket{\mathbf{k}_1}_1 
=\frac{1}{\Omega}\sum_{G_1,G_2}\delta_{\mathbf{k}_1+G_1,\mathbf{k}_2+G_2}t(\mathbf{k}_1+G_1)e^{iG_2 \cdot r_d},
\end{multline}
where $t(\mathbf{q})\equiv\int T(\mathbf{r}+d_0 \mathbf{e_z})e^{-i\mathbf{q}\cdot\mathbf{r}}d^2\mathbf{r}$, $\Omega$ is the area of the monolayer Wigner-Seitz cell, and $G_j$ is the reciprocal lattice vector of the $j$th layer. 
Eq.~(\ref{eq.generalumklapp}) show that the interlayer Hamiltonian couples Bloch states $\ket{\mathbf{k}_1(\mathbf{q})}_1$ and $\ket{\mathbf{k}_2(\mathbf{q})}_2$ 
satisfying the generalized Umklapp scattering condition, $\mathbf{k}_1+\mathbf{G}_1=\mathbf{k}_2+\mathbf{G}_2\equiv \mathbf{q}$. 
The coupling strength is proportional to $|t(\mathbf{q})|$ which decays exponentially as $|\mathbf{q}|$ increases,
so that the low energy properties of the Hamiltonian can be described by the momentum region with small $|\mathbf{q}|$.
This motivate us to introduce the quasiband model, composed of the Bloch states with small momentum within the effective BZ (EBZ).
Here we take the union of the first BZs (UBZ) of two monolayers as the EBZ,
and neglect the states with the momentum outside the EBZ.

Explicitly, the quasiband model $\mathbf{H}_{QB}=\sum_{\mathbf{q}\in\text{EBZ}}\mathbf{H(q)}$ 
with $\mathbf{H(q)}=\left[t(\mathbf{q})\ket{\mathbf{k}_1(\mathbf{q})}_1{}_2\bra{\mathbf{k}_2(\mathbf{q})}+h.c.\right]$
can be divided as $\mathbf{H}_{QB}=\bf{H_A}+\bf{H_B}+\bf{H_C}$ where $A$, $B$, $C$ denote three regions within the UBZ as shown in Fig.~\ref{fig:triangle}(a). 
First, in region A, $\ket{\bf{q}}_1$ can couple only to $\ket{\bf{q}}_2$ because any momentum of the form $\bf{q+G_{1,2}}$ 
is placed outside the EBZ. Thus $\bf{H_A(q)}$ in the basis of $(\ket{\bf{q}}_{1},\ket{\bf{q}}_{2})$ becomes 
\begin{align}
\label{eq.triangularHA}
\mathbf{H_A(q)} =
\left( \begin{array}{cc}
0 & T^t(\bf{q})^* \\
T^t(\bf{q}) & 0\\
\end{array} \right),
\end{align}
which gives two nonzero energy eigenvalues  $\pm|T^t(\bf{q})|$ where $T^t(\mathbf{q})=(-1/\Omega_t) t(\bf{q})$ and $\Omega_t=\sqrt{3}a_0^2/2$.

On the other hand, in regions B and C, some momentum $\bf{q+G_{1,2}}$ with nonzero $\bf{G_{1,2}}$ can be placed within the EBZ.
For example, in region B, $\bf{q}+\bf{G_2}$ can be located inside the first BZ of the layer 1 [see Fig.~\ref{fig:triangle}(a,b)]. 
Because of that, $\ket{\bf{q}}_2$ can couple to both $\ket{\bf{q}}_1$ and $\ket{\bf{q+G_2}}_1$ 
so that $\bf{H_B(q)}$ becomes a $3\times 3$ matrix as
\begin{align}
\label{eq.triangular3*3}
\mathbf{H_B(q)} =
\left( \begin{array}{ccc}
0 & T^t_1(q)^* & 0 \\
T^t_1(q) & 0 & T^t_2(q)^*\\
0 & T^t_2(q) & 0 \\
\end{array} \right),
\end{align}
which gives three eigenvalues 0 and $\pm\sqrt{|T^t_1(\mathbf{q})|^2 + |T^t_2(\mathbf{q})|^2}$ 
where $T^t_1(\mathbf{q})=(-1/\Omega_t) t(\mathbf{q})$, $T^t_2(\mathbf{q})=(-1/\Omega_t) t(\bf{q+G_2})$. 
We note that chiral symmetry guarantees the presence of one ZES for the odd-dimensional Hamiltonian $\bf{H_B(q)}$. 
The wavefunction of the ZES is $\ket{\psi^1_0(\mathbf{q})}\propto \left[T^t_2(q)^*\ket{\mathbf{q}}_1-T^t_1(q)\ket{\mathbf{q+G_2}}_1\right]$ 
localized in the layer 1. 
Therefore in region B, there always is at least one ZES per three coupled Bloch states. 
Same argument can also be applied to region C by switching the role of layer 1 and 2.

Fig.~\ref{fig:triangle}(c) compares DOS from the full tightbinding model and the quasiband model.
The low energy properties, such as the number of ZESs and the magnitude of the gap, match very well. 
To understand the influence of large momentum states on the ZESs, we have identified the momentum region which can couple the ZESs and compared the coupling strength between ZESs with the energy gap. 
As ZESs from region B (C) are located within the layer 1 (2) and the Hamiltonian contains the interlayer coupling only,
the coupling between ZESs is possible only in the momentum region (dark filled region (DFR) in Fig.~\ref{fig:triangle}(d)) 
where the region B of the layer 1 and region C of the layer 2 overlap in the extended BZ.  
Since DFRs are farther away from the EBZ and the relevant coupling strength is proportional to $|t(\bf{q})|$, decaying exponentially as $|\bf{q}|$ increases,
the energy splitting between the ZESs due to DFRs is much smaller than 
the energy gap that is determined by $|t(\bf{q}_{c})|$ with the momentum $\bf{q}_{c}$ at the EBZ boundary. 
Therefore the quasiband model correctly captures the low energy electronic properties.

Interestingly, using the quasiband model, one can easily estimate the ratio of the number of ZESs to the total number of states.
As a Bloch state can be assigned at every momentum, the total number of states is proportional to the area of EBZ. 
Since one ZES appears per 3 coupled Bloch states in region B and C, 
the total number of ZESs is proportional to the area of region B and C, which gives the number ratio $P_{\text{TBTL}}=(2-\sqrt{3})^2\simeq0.072$,
identical to numerical results.

%%%%%%%%%%%%%%%%%%%%%%%%%%%%%%%%%%%%%%%%   FIGURE   %%%%%%%%%%%%%%%%%%%%%%%%%%%%%%%%%%%%%%%%%%%%%%%%%%%%%%
\begin{figure}
\includegraphics[width=0.5\textwidth]{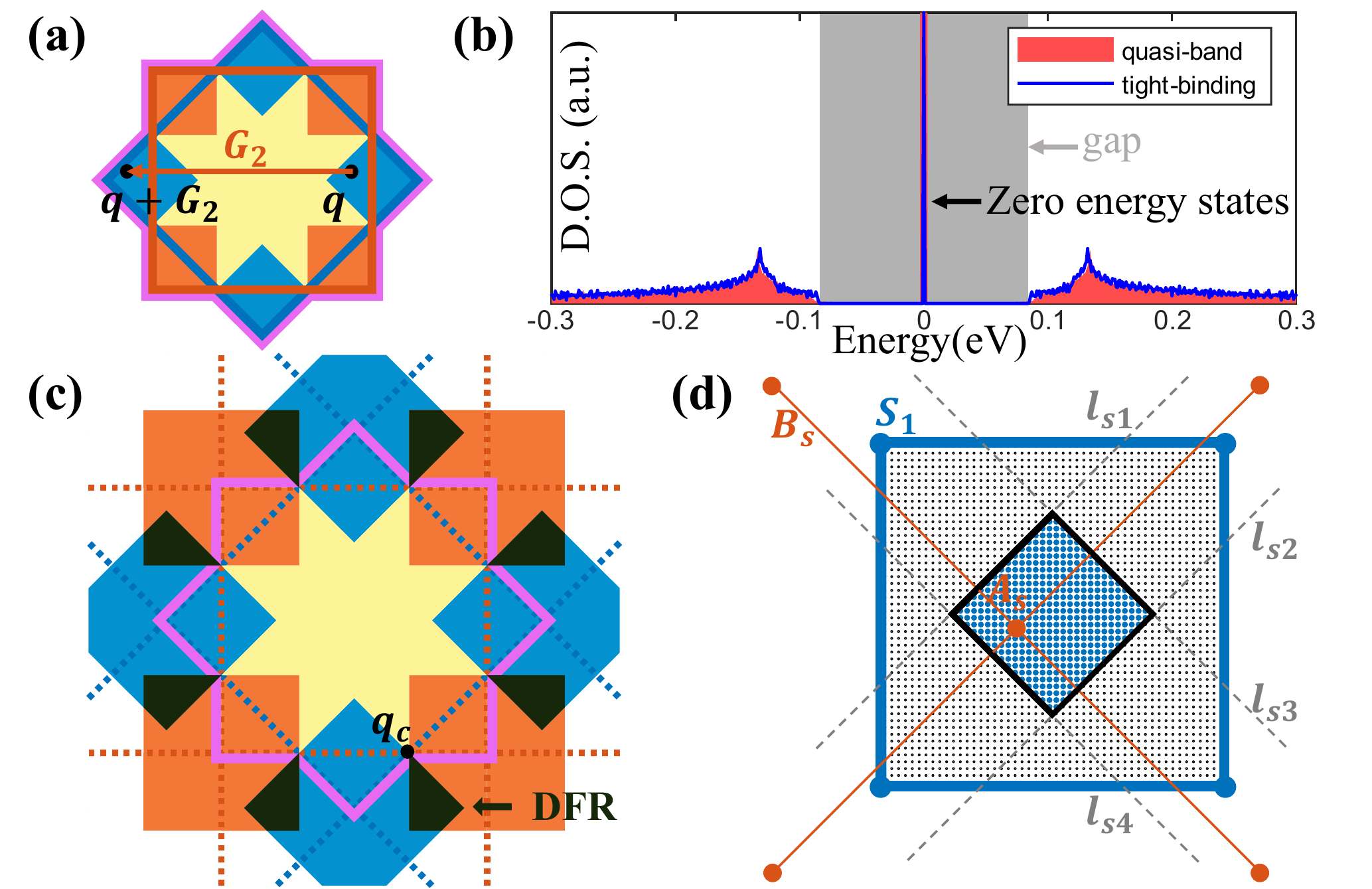}
\caption{
(a) The red and blue squares indicate the first BZs of the layer 1 and 2 whose union gives the EBZ (purple). 
EBZ is divided into three regions $A$ (yellow), $B$ (blue), $C$ (orange) as in Fig~\ref{fig:triangle}.
(b) DOS from the full tight-binding model and the quasiband model.
(c) Regions $B$ and $C$, and DFR illustrated in the extended BZ scheme.
(d) Real space geometric consideration for ZESs. Orange dots denote the layer 2 where ZESs are confined. 
The small square with blue dots inside indicates the possible center positions of ZESs with maximum amplitude. 
}
\label{fig:square}
\end{figure}
%%%%%%%%%%%%%%%%%%%%%%%%%%%%%%%%%%%%%%%%   FIGURE   %%%%%%%%%%%%%%%%%%%%%%%%%%%%%%%%%%%%%%%%%%%%%%%%%%%%%%

{\it Counting ZESs from real space geometry.---}
Linear superpositions of ZESs can form exponentially localized ZESs (ELZES) in real space.
Explicitly, one can find the same number of ELZESs and in-gap ZESs in all three types of bilayer quasicrystals (see SM). 
Interestingly, when chiral symmetry $U$ exists, an ELZES $\ket{L}$ can always be confined within a single layer. 
This is because, as $U$ changes the sign of the states in the layer 2, 
$\ket{L}+U\ket{L}$ ($\ket{L}-U\ket{L}$) is confined within the layer 1 (layer 2). 
Suppose that $\ket{L_1}$ is a ZES confined in the layer 1, given by
$\ket{L_1}=\sum_{i}a(\mathbf{R_{1i}})\ket{\mathbf{R_{1i}}}$.
From $\mathbf{H}\ket{L_1}=0$, we obtain
$\sum_{i,j}a(\mathbf{R_{1i}})T(\mathbf{R_{1i}}-\mathbf{R_{2j}})\ket{\mathbf{R_{2j}}}=0$, which gives
$\sum_{i}a(\mathbf{R_{1i}})T(\mathbf{R_{1i}}-\mathbf{R_{2j}})=0$ for any $\ket{\mathbf{R_{2j}}}$.
Namely, for a given site $\mathbf{R_{2j}}$ in the layer 2, when the relevant hopping amplitude to the site $\mathbf{R_{1i}}$ in the layer 1, weighted by $a(\mathbf{R_{1i}})$, is summed over all possible $i$, we obtain zero.
From this, one can derive a compact real-space geometric condition to count ELZESs as follows.

Let us consider an ELZES in the layer 2 whose maximum amplitude is at the site $A_t$. 
In Fig.~\ref{fig:triangle}(e), we plot a unit cell (blue triangle) in the layer 1 which embraces $A_t$ inside. 
We note that the total weighted hopping amplitude to the site $T_1$ in the layer 1 should be zero as shown above, 
and the major hopping amplitudes to $T_1$ come from the sites $B_t$ and $A_t$ in the layer 2. 
Since the wave function amplitude at $A_t$ should be bigger than that at $B_t$, and the hopping term gets smaller as the hopping distance is farther, 
the length of $\overline{B_t T_1}$ should be shorter than that of $\overline{A_t T_1}$.
Hence the location of $A_t$ is constrained to be placed under the line $l_{t1}$, which is perpendicular to $\overline{A_{t}B_{t}}$ and whose distance to $T_1$ 
is a half of the lattice constant. 
Similar consideration of the other two corners of the blue triangle shows that
$A_t$ can maintain the maximum amplitude of the ELZES when it is confined within a small triangle formed by three lines $l_{t1}$, $l_{t2}$, and $l_{t3}$ shown in Fig.~\ref{fig:triangle}(e).
Because of the aperiodicity, 
if one traces the relative positions of lattice sites of the layer 2 projected to the triangular unit cells of the layer 1, and collects all the projected positions
within a single triangular unitcell of the layer 1, one can observe uniform distribution of projected points as in Fig.~\ref{fig:triangle}(e). 
This means that we can estimate the number of ZESs with respect to the total number of unit cells in the system 
by evaluating the area of the small triangle defined by $l_{t1,t2,t3}$ over the area of the unit triangular cell, 
which is equal to $P_{tri}=(2-\sqrt{3})^2=0.072$. This is further confirmed numerically (see SM). 
Interestingly, this ratio is exactly identical to the estimation based on the quasiband method.

{\it Twisted bilayer square lattice (TBSL) quasicrystal.---}
It is straightforward to construct the strong coupling Hamiltonian for TBSL quasicrystal with $\theta=45\degree$ using Eq.~(\ref{eq.LatticeH}).
The relevant quasiband Hamiltonian can also be obtained by taking the relevant UBZ as the EBZ as shown in Fig.~\ref{fig:square}(a).
The quasiband model correctly describes the number of ZESs and the energy gap, consistent with the full tightbinding model (see Fig.~\ref{fig:square}(b)).
Also, the ratio of the ZES number over the total state number 
can be correctly estimated from the area of regions $B$, $C$, which gives $P_{\text{TBSL}}=3-2\sqrt{2}$.
The same $P_{\text{TBSL}}$ can be obtained from real space geometric condition for maximum amplitude positions of ZESs shown in Fig~\ref{fig:square}(c). Here $P_{\text{TBSL}}$ is given by the area of the small square with blue dots over the area of the blue square with gray dots.

%%%%%%%%%%%%%%%%%%%%%%%%%%%%%%%%%%%%%%%%   FIGURE   %%%%%%%%%%%%%%%%%%%%%%%%%%%%%%%%%%%%%%%%%%%%%%%%%%%%%%
\begin{figure}
\includegraphics[width=0.5\textwidth]{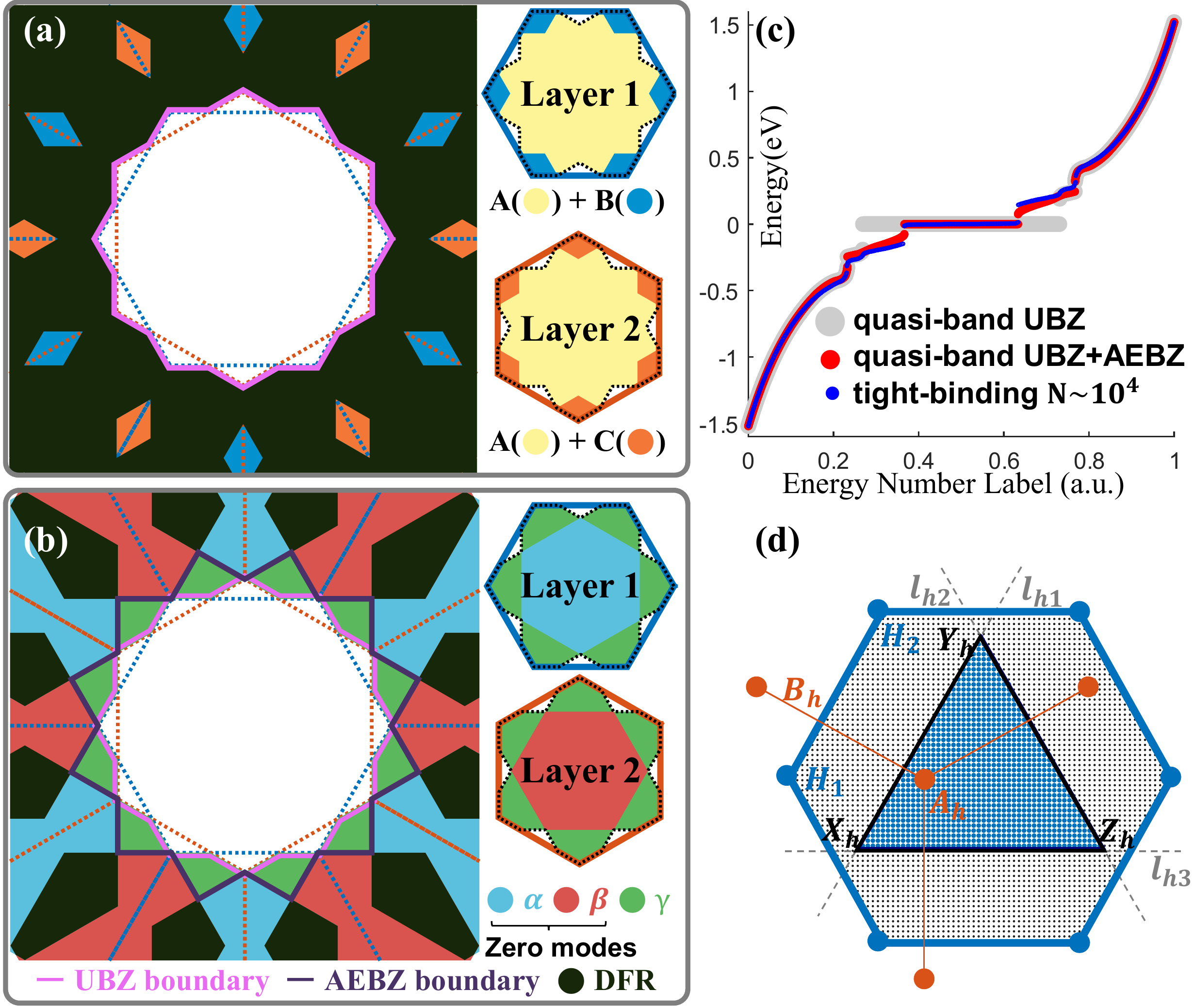}
\caption{
(a) Regions $B$, $C$, and DFR when the UBZ is taken as the EBZ whose boundary is shown in purple color.
EBZ is again divided into three regions $A$, $B$, $C$ as in Fig.~\ref{fig:triangle}(a).
(b) Regions $\alpha$, $\beta$, and DFR when the adjusted EBZ (AEBZ) is additionally taken for the quasiband model.
(c) Energy spectra from the full tight-binding model and the quasiband model with UBZ and (UBZ+AEBZ).
(d) Real space geometry for ZESs.
}
\label{fig:hexagon}
\end{figure}
%%%%%%%%%%%%%%%%%%%%%%%%%%%%%%%%%%%%%%%%   FIGURE   %%%%%%%%%%%%%%%%%%%%%%%%%%%%%%%%%%%%%%%%%%%%%%%%%%%%%%

{\it TBG quasicrystal.---} It is straightforward to apply the Hamiltonian in Eq.~(\ref{eq.LatticeH})
to TBG quasicrystal, taking $a_0=0.142~\mathrm{nm}$ as the distance between nearest-neighboring sites. 
The corresponding quasiband model can also be constructed by taking the UBZ as the EBZ, which is again divided into $A$, $B$, $C$ regions.
The only difference compared to TBTL quasicrystal is that there are two eigenstates per each momentum $\bf{q}$
in the BZ because of the sublattice degrees of freedom, ($a$, $b$), in the honeycomb lattice.

Therefore, the Hamiltonian $\bf{H_A(q)}$ for region $A$ becomes a $4\times4$ matrix and $\bf{H_B(q)}$, $\bf{H_C(q)}$ become $6\times6$ matrices.
From diagonalizing the matrices, one can find ZESs for layer 1 in region A and B and ZESs for layer 2 in region A and C as described in the right panel of Fig.~\ref{fig:hexagon}(a). All the ZESs have the form $(1/\sqrt{2})(\ket{\mathbf{q},a}_i-\ket{\mathbf{q},b}_i)$ where a, b denote the sublattices [see SM for details].

The existence of ZESs is guaranteed by the chiral symmetry $U$ together with the sublattice exchange symmetry $U_{ab}^i(\mathbf{q})$  acting at $\mathbf{q}$, which emerges in the quasiband model
because of the momentum cutoff of the EBZ (see SM).
Explicitly, $U_{ab}^i(\mathbf{q})=\ket{\mathbf{q},a}_i{}_i\bra{\mathbf{q},b} + \ket{\mathbf{q},b}_i{}_i\bra{\mathbf{q},a}$ where $i=1,~2$ is the layer index,
which satisfies $[U,U_{ab}^i(\mathbf{q})]=0$ for any $i$ and $\mathbf{q}$, and two $U_{ab}^i(\mathbf{q})$ with different $i$ or $\mathbf{q}$ commute.
Also, as $\left[U_{ab}^i(\mathbf{q})\right]^2=I$, its eigenvalues are $\pm1$. 
In each region, the quasiband Hamiltonian can be block-diagonalized with blocks having different $U_{ab}^i$ eigenvalues. The presence of blocks with odd-dimensions guarantees the ZESs [see SM.]

In Fig.~\ref{fig:hexagon}(c), we compare the energy spectra of the full lattice model and the quasiband model with UBZ as EBZ.
Contrary to the cases of TBTL and TBSL quasicrystals, these two models give inconsistent results.
The discrepancy can be understood in two different ways.
One is from the fact that $U_{ab}^i(\mathbf{q})$ is not a physical symmetry but an emergent symmetry
owing to the presence of the momentum cutoff in the EBZ.
Thus $U_{ab}^i(\mathbf{q})$ is broken and some of predicted ZESs are gapped when the region outside the UBZ is included.
Second is due to the large coupling strength between predicted ZESs arising from the DFR, plotted in Fig.~\ref{fig:hexagon}(a).
Contrary to TBTL and TBSL quasicrystals, 
the DFR overlap along the full boundary of the EBZ in TBG quasicrystals.
We find that the coupling strength between ZESs is comparable to the energy gap so that the quasiband model
is unreliable.

To remedy this problem, we take into account the large momentum region outside UBZ, referred to as the adjusted EBZ (AEBZ) in Fig.4. Here the AEBZ is chosen so that the DFR for the new EBZ including the UBZ and AEBZ, becomes apart from the AEBZ. Then using the extended quasiband model, we examine the influence of the AEBZ on ZESs. 
A systematic way of constructing AEBZ is explained in SM.

Fig.~\ref{fig:hexagon}(b) shows the new DFR, which touches the boundary of AEBZ only at few points as in TBSL quasicrystal.
In this AEBZ, two states from different layers, each belongs to region $\gamma$, can be coupled while states from region $\alpha$ from layer 1 and region $\beta$ from layer 2 remains as ZESs.
Fig.~\ref{fig:hexagon}(c) compares the energy spectra from the quasiband model with AEBZ and the full lattice models,
which match very well.
We have confirmed that including states outside AEBZ further does not change the number of ZESs and the AEBZ successfully explain
the low energy electronic structures [see SM]. 

As before, one can estimate the ratio of the ZES number to the total state number
by comparing the area of the patches in the extended EBZ, 
whch is given by $P_{\text{TBG}}=2-\sqrt{3}\simeq0.268$.
The same $P_{\text{TBG}}$ can be obtained from the real space geometric consideration of the maximum amplitude of ELZESs (see Fig.~\ref{fig:hexagon}(d)).

{\it Conclusion.---}
We showed that the TBTL, TBSL, TBG quasicrystals in strong coupling limit generally support macroscopically degenerate ZESs.
The interplay of the aperiodicity and the emergent chiral symmetry plays a critical role on the presence of the ZESs, and their spatial localization,
and their counting.
We further confirmed that as long as the intralayer hopping amplitudes are smaller than the energy gap in the strong couplimg limit,
the ZESs still appear as ingap states, though they are not exactly at zero energy.
To observe the strong coupling physics of bilayer quasicrystals, one can either apply strong vertical pressure or make artifical lattice
structures using metamaterials.
We believe that bilayer quasicrystals are a promising avenue to observe the intriguing interplay of lattice aperiodicity and flat band physics.

\begin{acknowledgments}
We thank Jae-Mo Lihm and Yoonseok Hwang for helpful discussions.  
H.H. was supported by Samsung  Science and Technology Foundation under Project Number SSTF-BA2002-06.
B.J.Y. was supported by the Institute for Basic Science in Korea (Grant No. IBS-R009-D1), 
Samsung  Science and Technology Foundation under Project Number SSTF-BA2002-06,
Basic Science Research Program through
the National Research Foundation of Korea (NRF) (Grant No. 0426-20200003).
This work was supported in part by the U.S. Army Research Office and and Asian Office of Aerospace Research \& Development (AOARD) under Grant Number W911NF-18-1-0137.
\end{acknowledgments}

%%%%%%%%%%%%%%%%%%%%%%%%%%%%%%%%%%%%%%%%%%%%%%%%%%%%%%%%%%%%%%%%%%%%%%%%%%%
%\bibliographystyle{unsrt} 
%\bibliography{main}

\begin{thebibliography}{100}

\bibitem{flat_highTsc}
Tero~T Heikkil{\"a} and Grigory~E Volovik.
\newblock Flat bands as a route to high-temperature superconductivity in
  graphite.
\newblock In {\em Basic Physics of Functionalized Graphite}, pages 123--143.
  Springer, 2016.

\bibitem{flat_FQHE1}
Evelyn Tang, Jia-Wei Mei, and Xiao-Gang Wen.
\newblock High-temperature fractional quantum hall states.
\newblock {\em Phys. Rev. Lett.}, 106:236802, Jun 2011.

\bibitem{flat_FQHE2}
Kai Sun, Zhengcheng Gu, Hosho Katsura, and S.~Das~Sarma.
\newblock Nearly flatbands with nontrivial topology.
\newblock {\em Phys. Rev. Lett.}, 106:236803, Jun 2011.

\bibitem{flat_FQHE3}
Titus Neupert, Luiz Santos, Claudio Chamon, and Christopher Mudry.
\newblock Fractional quantum hall states at zero magnetic field.
\newblock {\em Phys. Rev. Lett.}, 106:236804, Jun 2011.

\bibitem{flat_FQHE4}
N.~Regnault and B.~Andrei Bernevig.
\newblock Fractional chern insulator.
\newblock {\em Phys. Rev. X}, 1:021014, Dec 2011.

\bibitem{flat_FQHE5}
DN~Sheng, Zheng-Cheng Gu, Kai Sun, and L~Sheng.
\newblock Fractional quantum hall effect in the absence of landau levels.
\newblock {\em Nature communications}, 2(1):1--5, 2011.

\bibitem{flat_Wigner}
Congjun Wu, Doron Bergman, Leon Balents, and S.~Das~Sarma.
\newblock Flat bands and wigner crystallization in the honeycomb optical
  lattice.
\newblock {\em Phys. Rev. Lett.}, 99:070401, Aug 2007.

\bibitem{magicangle}
Rafi Bistritzer and Allan~H MacDonald.
\newblock Moir{\'e} bands in twisted double-layer graphene.
\newblock {\em Proceedings of the National Academy of Sciences},
  108(30):12233--12237, 2011.

\bibitem{CastroNeto_TBGelectronicstructure}
JMB~Lopes Dos~Santos, NMR Peres, and AH~Castro Neto.
\newblock Graphene bilayer with a twist: Electronic structure.
\newblock {\em Physical review letters}, 99(25):256802, 2007.

\bibitem{CastroNeto_TBGcontinuum}
JMB~Lopes Dos~Santos, NMR Peres, and AH~Castro Neto.
\newblock Continuum model of the twisted graphene bilayer.
\newblock {\em Physical Review B}, 86(15):155449, 2012.

\bibitem{Kindermann_continuum}
M~Kindermann and PN~First.
\newblock Local sublattice-symmetry breaking in rotationally faulted multilayer
  graphene.
\newblock {\em Physical Review B}, 83(4):045425, 2011.

\bibitem{koshino_moon_optical}
Pilkyung Moon and Mikito Koshino.
\newblock Optical absorption in twisted bilayer graphene.
\newblock {\em Physical Review B}, 87(20):205404, 2013.

\bibitem{tbg_PJHmott}
Yuan Cao, Valla Fatemi, Ahmet Demir, Shiang Fang, Spencer~L Tomarken, Jason~Y
  Luo, Javier~D Sanchez-Yamagishi, Kenji Watanabe, Takashi Taniguchi, Efthimios
  Kaxiras, et~al.
\newblock Correlated insulator behaviour at half-filling in magic-angle
  graphene superlattices.
\newblock {\em Nature}, 556(7699):80--84, 2018.

\bibitem{tbg_PJHsc}
Yuan Cao, Valla Fatemi, Shiang Fang, Kenji Watanabe, Takashi Taniguchi,
  Efthimios Kaxiras, and Pablo Jarillo-Herrero.
\newblock Unconventional superconductivity in magic-angle graphene
  superlattices.
\newblock {\em Nature}, 556(7699):43--50, 2018.

\bibitem{tbg_sc}
Matthew Yankowitz, Shaowen Chen, Hryhoriy Polshyn, Yuxuan Zhang, K~Watanabe,
  T~Taniguchi, David Graf, Andrea~F Young, and Cory~R Dean.
\newblock Tuning superconductivity in twisted bilayer graphene.
\newblock {\em Science}, 363(6431):1059--1064, 2019.

\bibitem{no_translational_sym1}
Dan Shechtman, Ilan Blech, Denis Gratias, and John~W Cahn.
\newblock Metallic phase with long-range orientational order and no
  translational symmetry.
\newblock {\em Physical review letters}, 53(20):1951, 1984.

\bibitem{no_translational_sym2}
AI~Goldman and M~Widom.
\newblock Quasicrystal structure and properties.
\newblock {\em Annual Review of Physical Chemistry}, 42(1):685--729, 1991.

\bibitem{stampfli}
P~Stampfli.
\newblock A dodecagonal quasi-periodic lattice in 2 dimensions.
\newblock {\em Helvetica Physica Acta}, 59(6-7):1260--1263, 1986.

\bibitem{stampfli2}
Elad Koren and Urs Duerig.
\newblock Superlubricity in quasicrystalline twisted bilayer graphene.
\newblock {\em Physical Review B}, 93(20):201404, 2016.

\bibitem{ahn}
Sung~Joon Ahn, Pilkyung Moon, Tae-Hoon Kim, Hyun-Woo Kim, Ha-Chul Shin, Eun~Hye
  Kim, Hyun~Woo Cha, Se-Jong Kahng, Philip Kim, Mikito Koshino, et~al.
\newblock Dirac electrons in a dodecagonal graphene quasicrystal.
\newblock {\em Science}, 361(6404):782--786, 2018.

\bibitem{Son}
Pilkyung Moon, Mikito Koshino, and Young-Woo Son.
\newblock Quasicrystalline electronic states in 30 rotated twisted bilayer
  graphene.
\newblock {\em Physical Review B}, 99(16):165430, 2019.

\bibitem{Sungbin}
Moon~Jip Park, Hee~Seung Kim, and SungBin Lee.
\newblock Emergent localization in dodecagonal bilayer quasicrystals.
\newblock {\em Physical Review B}, 99(24):245401, 2019.

\bibitem{pseudogap}
Takeo Fujiwara and Takeshi Yokokawa.
\newblock Universal pseudogap at fermi energy in quasicrystals.
\newblock {\em Physical review letters}, 66(3):333, 1991.

\bibitem{pseudogap_NMR}
X-P Tang, EA~Hill, SK~Wonnell, SJ~Poon, and Y~Wu.
\newblock Sharp feature in the pseudogap of quasicrystals detected by nmr.
\newblock {\em Physical review letters}, 79(6):1070, 1997.

\bibitem{pseudogap_Xray}
K~Kirihara, T~Nagata, K~Kimura, K~Kato, M~Takata, E~Nishibori, and M~Sakata.
\newblock Covalent bonds and their crucial effects on pseudogap formation in
  $\alpha$- al (m n, r e) si icosahedral quasicrystalline approximant.
\newblock {\em Physical Review B}, 68(1):014205, 2003.

\bibitem{penrose}
Roger Penrose.
\newblock The role of aesthetics in pure and applied mathematical research.
\newblock {\em Bull. Inst. Math. Appl.}, 10:266--271, 1974.

\bibitem{penrose_ratio}
Ezra Day-Roberts, Rafael~M Fernandes, and Alex Kamenev.
\newblock Nature of protected zero-energy states in penrose quasicrystals.
\newblock {\em Physical Review B}, 102(6):064210, 2020.

\bibitem{penrose_perp}
Murod Mirzhalilov and M.~\"O. Oktel.
\newblock Perpendicular space accounting of localized states in a quasicrystal.
\newblock {\em Phys. Rev. B}, 102:064213, Aug 2020.

\bibitem{penrose_localize}
Masao Arai, Tetsuji Tokihiro, Takeo Fujiwara, and Mahito Kohmoto.
\newblock Strictly localized states on a two-dimensional penrose lattice.
\newblock {\em Physical Review B}, 38(3):1621, 1988.

\bibitem{penrose_E}
Mahito Kohmoto and Bill Sutherland.
\newblock Electronic states on a penrose lattice.
\newblock {\em Physical review letters}, 56(25):2740, 1986.

\bibitem{penrose_E2}
T~Odagaki and Dan Nguyen.
\newblock Electronic and vibrational spectra of two-dimensional quasicrystals.
\newblock {\em Physical Review B}, 33(4):2184, 1986.

\bibitem{penrose_E3}
TC~Choy.
\newblock Density of states for a two-dimensional penrose lattice: Evidence of
  a strong van-hove singularity.
\newblock {\em Physical review letters}, 55(26):2915, 1985.

\bibitem{penrose_hubbard}
Akihisa Koga and Hirokazu Tsunetsugu.
\newblock Antiferromagnetic order in the hubbard model on the penrose lattice.
\newblock {\em Physical Review B}, 96(21):214402, 2017.

\bibitem{maximallylocalizedwannier}
Nicola Marzari and David Vanderbilt.
\newblock Maximally localized generalized wannier functions for composite
  energy bands.
\newblock {\em Physical review B}, 56(20):12847, 1997.

\bibitem{graphenetransfer}
G~Trambly~de Laissardi{\`e}re, Didier Mayou, and Laurence Magaud.
\newblock Localization of dirac electrons in rotated graphene bilayers.
\newblock {\em Nano letters}, 10(3):804--808, 2010.

\bibitem{hofstadter}
Pilkyung Moon and Mikito Koshino.
\newblock Optical properties of the hofstadter butterfly in the moir{\'e}
  superlattice.
\newblock {\em Physical Review B}, 88(24):241412, 2013.

\bibitem{meleTBG}
Eugene~J Mele.
\newblock Commensuration and interlayer coherence in twisted bilayer graphene.
\newblock {\em Physical Review B}, 81(16):161405, 2010.

\bibitem{koshinointerlayer}
Mikito Koshino.
\newblock Interlayer interaction in general incommensurate atomic layers.
\newblock {\em New Journal of Physics}, 17(1):015014, 2015.

\bibitem{starpolygon}
Eric~W Weisstein.
\newblock Star polygon.
\newblock {\em https://mathworld. wolfram. com/}, 2010.



\bibitem{slaterkoster}
John~C Slater and George~F Koster.
\newblock Simplified lcao method for the periodic potential problem.
\newblock {\em Physical Review}, 94(6):1498, 1954.


\end{thebibliography}

\end{document}

%\begin{comment}
%%%%%%%%%%%%%TITLE%%%%%%%%%%%%%
\title{Supplemental Material: Macroscopically degenerate localized zero-energy states of quasicrystalline bilayer systems in strong coupling limit}
%%%%%%%%%%%%%%%%%%%%%%%%%%%%%%%

%%%%%%%%%%%%AUTHORS%%%%%%%%%%%%
\author{Hyunsoo \surname{Ha}}
\affiliation{Department of Physics and Astronomy, Seoul National University, Seoul 08826, Korea}

\author{Bohm-Jung \surname{Yang}}
\email{bjyang@snu.ac.kr}
\affiliation{Center for Correlated Electron Systems, Institute for Basic Science (IBS), Seoul 08826, Korea}
\affiliation{Department of Physics and Astronomy, Seoul National University, Seoul 08826, Korea}
\affiliation{Center for Theoretical Physics (CTP), Seoul National University, Seoul 08826, Korea}
%%%%%%%%%%%%%%%%%%%%%%%%%%%%%%%
%\end{comment}
%%%%%%%%%%%%%%%%%%%
%%%%%%%%%%%%%%%%%%%
%%%%%%%%%%%%%%%%%%%

\maketitle
%%%%%%%%%%%%%%%%%%%%%%%%%%%%%%%
%%%%%%%%%%%%%%%%%%%%%%%%%%%%%%%
\clearpage
\onecolumngrid

%\begin{abstract}
In this supplemental material, we (i) provide details in twisted bilayer triangular lattice and the effect of the interaction range $r_{\star}$ in the electronic structure near zero energy, (ii) introduce the numerical calculation for constructing the ELZESs and verifying the geometric conditions for ELZESs, (iii) provide details in Twisted Bilayer Graphene and the concept of AEBZ, (iv) derive the energy spectrum is irrelevant to the displacement between two layers, and (v) explain how we considered the edge states originating from the finite-size effect.
%\end{abstract}

%%%%%%%%%%%%%%%%%%%%%%%%%%%%%%%
\setcounter{section}{0}
\setcounter{figure}{0}
\setcounter{equation}{0}
\renewcommand{\thefigure}{S\arabic{figure}}
\renewcommand{\theequation}{S\arabic{equation}}
\renewcommand{\thesection}{S\arabic{section}}
\tableofcontents
%%%%%%%%%%%%%%%%%%%%%%%%%%%%%%%
\hfill \\
%\twocolumngrid
%%%%%%%%%%%%%%%%%%%%%%%%%%%%%%%

%%%%%%%%%%%%%%%%%%%%%%%%%%%%%%%%%%%%%%%%%%%%%%%%%%%%%%%%%%%%%%%%%%%%%%%%%%%%%%%%%%%%%%%%%%%%%%%%%%%%%%%%%%%%%%%%%%%%%%%%%%%%%%%%%%%%%%%%%%%%%%%%%%%%%%%%%%%%%%%%%%%%%%%%%%%%%%%%%%%%%%%%%%%%%%%%%%%%%%%%%%%%%%%%%%%%%%%%%%%%%%%%%%%%%%%%%%%%%%%%%%%%%%%%%%%%%%%%%%%%%%%%%%%%%%%%%%%%%%%%%%%%%%%%%%%%%%%%%%%%%%%%%%%%%%%%%%%%%%%%%%%%%%%%
\section{Twisted bilayer triangular lattice (TBTL) quasicrystal}
\label{section.triangular}
\subsection{Model Hamiltonian}

Let us first study a model for bilayer triangular lattices with $30\degree$ twist.
We place two triangular lattices, each with the in-plane ($xy$-plane) lattice constant $a_0$, on top of each other with the vertical ($z$-directional) distance $d_0$.
When the two layers are exactly aligned, the twist angle is defined to be zero.
To make a twist by an angle $\theta$, the top layer (layer 2) is rotated by $\theta$ counter-clockwisely about the $z$-axis with respect to a lattice site
while the bottom layer (layer 1) stays fixed.
We focus on the case when $\theta=30\degree$ at which the bilayer system loses the lattice translation symmetry but instead develops a quasi-crystalline structure.
After the rotation, the top layer can be displaced by $\mathbf{r_d}$ in the $xy$ plane with respect to the bottom layer.
In this work, we consider only the case with $\mathbf{r_d}=0$ since the electronic structure is independent with the displacement $r_{d}$ in the thermodynamic limit which is proven in ~\ref{displacement}.

For the bilayer with $30\degree$ twist, the lattice vector $\mathbf{R}_{i}$ for the layer $i=1,2$ can be written as
\begin{align}
    \label{eq.latticetriangular}
    \mathbf{R}_{1}&=n\mathbf{b}_1+m\mathbf{b}_2, \nonumber\\
    \mathbf{R}_{2}&=n'\mathbf{b'}_1+m'\mathbf{b'}_2+\mathbf{d_z}+\mathbf{r_d},
\end{align}
where $\mathbf{d_z}=(0,0,d_0)$. We set the Bravais lattice vectors of the layer 1 as $\mathbf{b}_1=(a_0,0,0)$, $\mathbf{b}_2=(a_0/2,\sqrt{3}a_0/2,0)$, and those of the layer 2 as $\mathbf{b'}_1=R_{z,\pi/6}\mathbf{b}_1$, $\mathbf{b'}_2=R_{z,\pi/6}\mathbf{b}_2$  where $R_{z,\theta}$ indicates the matrix representation for the rotation by $\theta$ counter-clockwisely about the $z$-axis.

We consider the Hamiltonian
\begin{align}
\label{eq.LatticeH}
H=-\sum_{\mathbf{R}_{1},\mathbf{R}_{2}}T(\mathbf{R}_{1}-\mathbf{R}_{2})\ket{\mathbf{R}_{1}}\bra{\mathbf{R}_{2}} +h.c.
\end{align}
where$\ket{\mathbf{R}_{1}}$ ($\ket{\mathbf{R}_{2}}$) indicates the $p_z$ orbital located at the position $\mathbf{R}_{1}$ ($\mathbf{R}_{2}$) in the layer 1 (layer 2).
$T(\mathbf{R}_{1}-\mathbf{R}_{2})$ is the transfer integral between two orbitals at the sites $\mathbf{R}_{1}$ and $\mathbf{R}_{2}$,
which is assumed to take the following form,
\begin{align}
\label{eq.transfer integral}
-T(\mathbf{R})&=V_{pp\pi}\left[1-\left(\frac{\mathbf{R}\cdot\mathbf{e}_z}{R}\right)^2\right]+V_{pp\sigma}\left(\frac{\mathbf{R}\cdot\mathbf{e}_z}{R}\right)^2,\nonumber\\
V_{pp\pi}&=V_{pp\pi}^0 e^{(-R-a_0)/r_{\star}}, V_{pp\sigma}=V_{pp\sigma}^0 e^{(-R-d_0)/r_{\star}}
\end{align}
where $\mathbf{R}$ is a relative displacement vector between two atoms, $R=|\mathbf{R}|$, $\mathbf{e}_z$ denotes the unit vector in the $z$-direction.
$V_{pp\pi}$ ($V_{pp\sigma}$) indicates the transfer integral between nearest-neighbor $p_z$ orbitals forming $\pi$-bonding ($\sigma$-bonding). 
The length scale $r_\star$ determines the range of hopping transfer. 
Taking into account the parameters of twisted bilayer graphene, we assume $a_0\approx0.142,\mathrm{nm}$, $d_0\approx0.335\,\mathrm{nm}$, $r_\star=r_0\approx 0.0453\,\mathrm{nm}$, $V_{pp\pi}^0\approx-2.6\,\mathrm{eV}$, and $V_{pp\sigma}^0\approx0.48 \,\mathrm{eV}$, as explained in the main text.

One intersting property of the Hamiltonian in Eq.~(\ref{eq.LatticeH}) is that it has chiral symmetry because only the interlayer hopping is allowed. 
Explicitly, for the following unitary operator
\begin{align}
    \label{eq.chiral}
    U=\sum_{\mathbf{R_{1}}}-\ket{R_{1}}\bra{R_{1}}+\sum_{\mathbf{R_{2}}}\ket{R_{2}}\bra{R_{2}},
\end{align}
one can easily show that $\{H,U\}=0$. Hence, for a given energy eigenstate $\ket{\psi}$ satisfying $H\ket{\psi}=E\ket{\psi}$, 
one can always find another energy eigenstate $U\ket{\psi}$ with the energy $-E$ so that the energy spectrum of the Hamiltonian
is symmetric with respect to the zero energy.

\subsection{Investigating the effect of $r_{\star}$ with the Quasiband Model}

%%%%%%%%%%%%%%%%%%%%%%%%%%%%%%%%%%%%%%%%%%%%%%%%%%%%%%
%FIGURE%
%%%%%%%%%%%%%%%%%%%%%%%%%%%%%%%%%%%%%%%%%%%%%%%%%%%%%%
\begin{figure}
\includegraphics[width=1.0\textwidth]{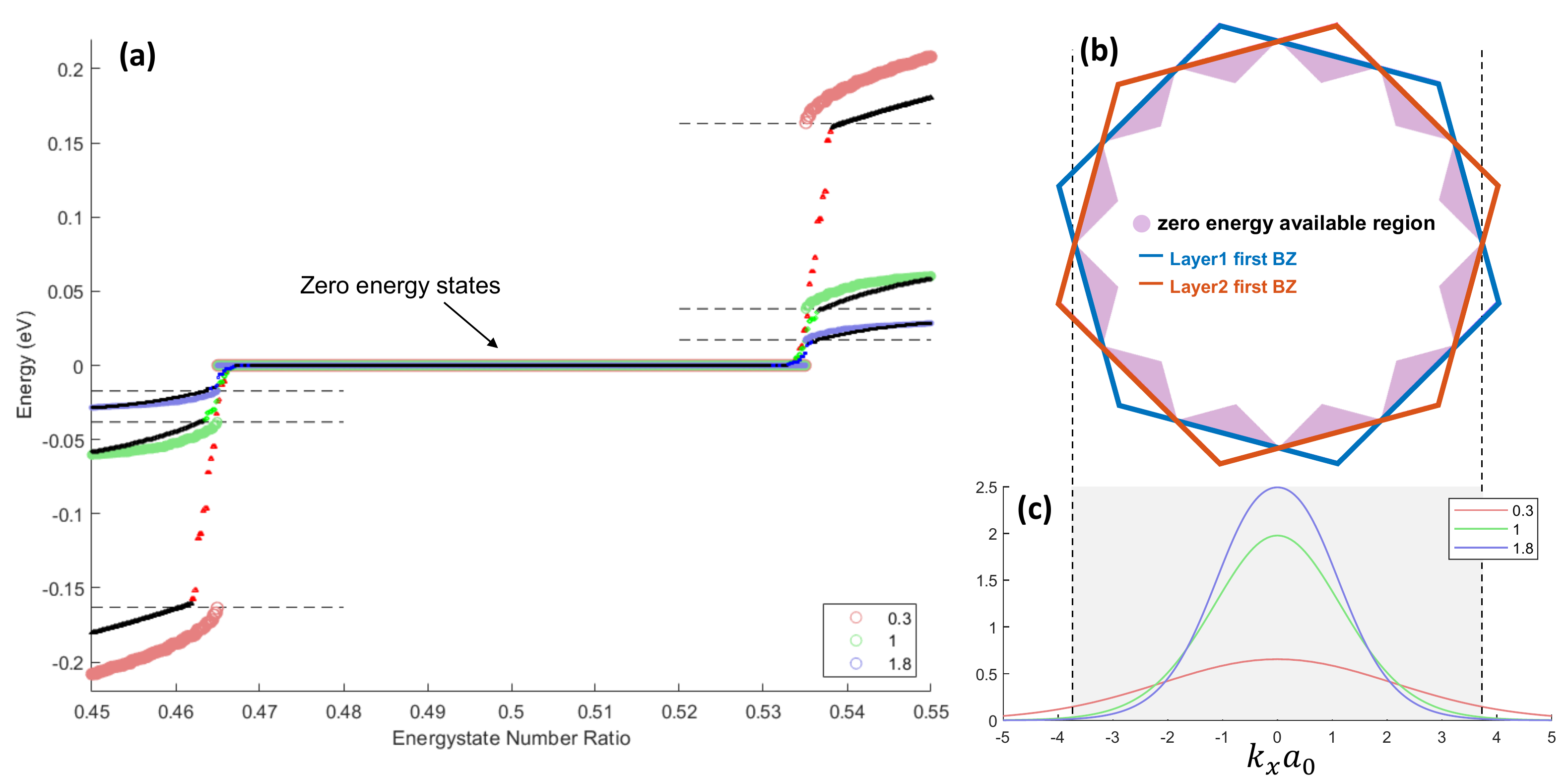}
\caption{
(a) Energy spectra near zero energy from the quasiband model and the tight binding model for various interaction ranges $r_\star/r_0$. 
Tight binding model calculations are performed for a finite-size system including about 20000 sites.
The colored dots indicate the results from the quasiband model with $r_\star/r_0=0.3$ (red), 1 (green), 1.8 (blue), respectively. 
The triangle, diamond, square symbols denote the data from the tight-binding model for $r_\star/r_0=0.3$, 1, 1.8, respectively. %
Unwanted edge modes arising from finite size effect are also highlighted with colors. 
Detailed procedure for determining the edge states is explained in Appendix \ref{edge}. 
(b) The momentum space area which contributes to zero energy states. 
(c) Profile of the Fourier components of interlayer hopping $|t(\mathbf{q})|$ for $r_\star/r_0=0.3$, 1, 1.8, respectively.
}
\label{fig.SFIG1}
\end{figure}
%%%%%%%%%%%%%%%%%%%%%%%%%%%%%%%%%%%%%%%%%%%%%%%%%%%%%%
%FIGURE%
%%%%%%%%%%%%%%%%%%%%%%%%%%%%%%%%%%%%%%%%%%%%%%%%%%%%%%

The strength of the mixing between Bloch states is proportional to $|t(\mathbf{q})|$ as described in the main text. To note with, $|t(\mathbf{q})|$ decays rapidly as $|\mathbf{q}|$ increases, and as $r_{\star}/r_0$ becomes larger, $|t(\bf{q})|/|t(0)|$ decays faster. 
For example, when $r_{\star}/r_0 = O(1)$, $|t(\bf{q}_{\text{out}})|/|t(\bf{q}=0)|$ with $\bf{q}_{\text{out}}$ outside the first BZ of each triangular layer is much smaller than $10^{-2}$, thus negligible.
Also, as $r_{\star}/r_0$ becomes larger, $|t(\bf{q})|/|t(0)|$ decays faster as $|\bf{q}|$ increases.

Considering the rapid decay of $|t(\mathbf{q})|$ as $|\mathbf{q}|$ grows, we introduce an effective BZ (EBZ) that is defined as the union of the two first BZs of two triangular layers rotated by $30\degree$,
and take the boundary of the EBZ as a momentum cutoff.
Then we predominantly consider the coupling of Bloch states with the momentum $\bf{q}$ inside the EBZ and neglect the Bloch states with $\bf{q}$ outside the EBZ.
The effective Hamiltonian composed of the Bloch states within the EBZ is called the quasiband model which can successsfully explain the number of zero energy modes
and the size of the gap above and below the zero energy, as shown in Fig.~\ref{fig.SFIG1}.
Furthermore, for $r_{\star}/r_0 \gtrsim 1$, the quasiband model successfully reproduce the electronic structure of the full lattice Hamiltonian as shown in the main text comparing the density of states between the quasiband model and the real lattice tight-binding calculation.

Fig.~\ref{fig.SFIG1}(a) shows the energy spectra near the zero energy from the tight-binding model and the quasiband model for various $r_{\star}/r_0$. 
One can find that the results from the two models, such as the apperance of degenerate zero energy states and the magnitude of the gap above and below the zero energy, are consistent. 
Comparing the energy from region A, B, and C (described in the main text), one can see that the band width is proportional to $|t(\mathbf{q}=0)|$ while the size of the gap above and below the zero energy is proportional to 
$|t(\mathbf{q})|$ with $\mathbf{q}$ at the boundary of the region A.
Hence when $r_{\star}/r_0$ is smaller, $|t(\mathbf{q})|$ spreads wider (Fig.~\ref{fig.SFIG1}(c)) and the gap size becomes bigger Fig.~\ref{fig.SFIG1}(a).
Although the contribution from the region outside the EBZ gives an error, the quasiband model gives satisfactory results, especially when $r_{\star}/r_0\gtrsim1$.
For example, density of states from the quasiband model and the tight-binding model are consistent when $r_{\star}/r_0=1$ as explained in the main text.

Interestingly, using the quasiband model, one can easily estimate the ratio of the number of zero energy states to the total number of states.
As a single point in the BZ of each layer is mapped to one Bloch state, the total number of states is proportional to the total area of the two first BZs. 
Considering that no zero energy state appears from region A while only one zero energy state appears per 3 coupled Bloch states in region B and C, 
the total number of zero energy states is proportional to the filled area in Fig.~\ref{fig.SFIG1}(b). 
This idea immediately shows that the number ratio for the bilayer triangular lattice should be $P_{tri}=(2-\sqrt{3})^2\simeq0.072$.

\section{Real Space Numerical Calculation of TBTL}
\label{section.realspace}

%%%%%%%%%%%%%%%%%%%%%%%%%%%%%%%%%%%%%%%%%%%%%%%%%%%%%%
%FIGURE%
%%%%%%%%%%%%%%%%%%%%%%%%%%%%%%%%%%%%%%%%%%%%%%%%%%%%%%
\begin{figure*}
\includegraphics[width=1.0\textwidth]{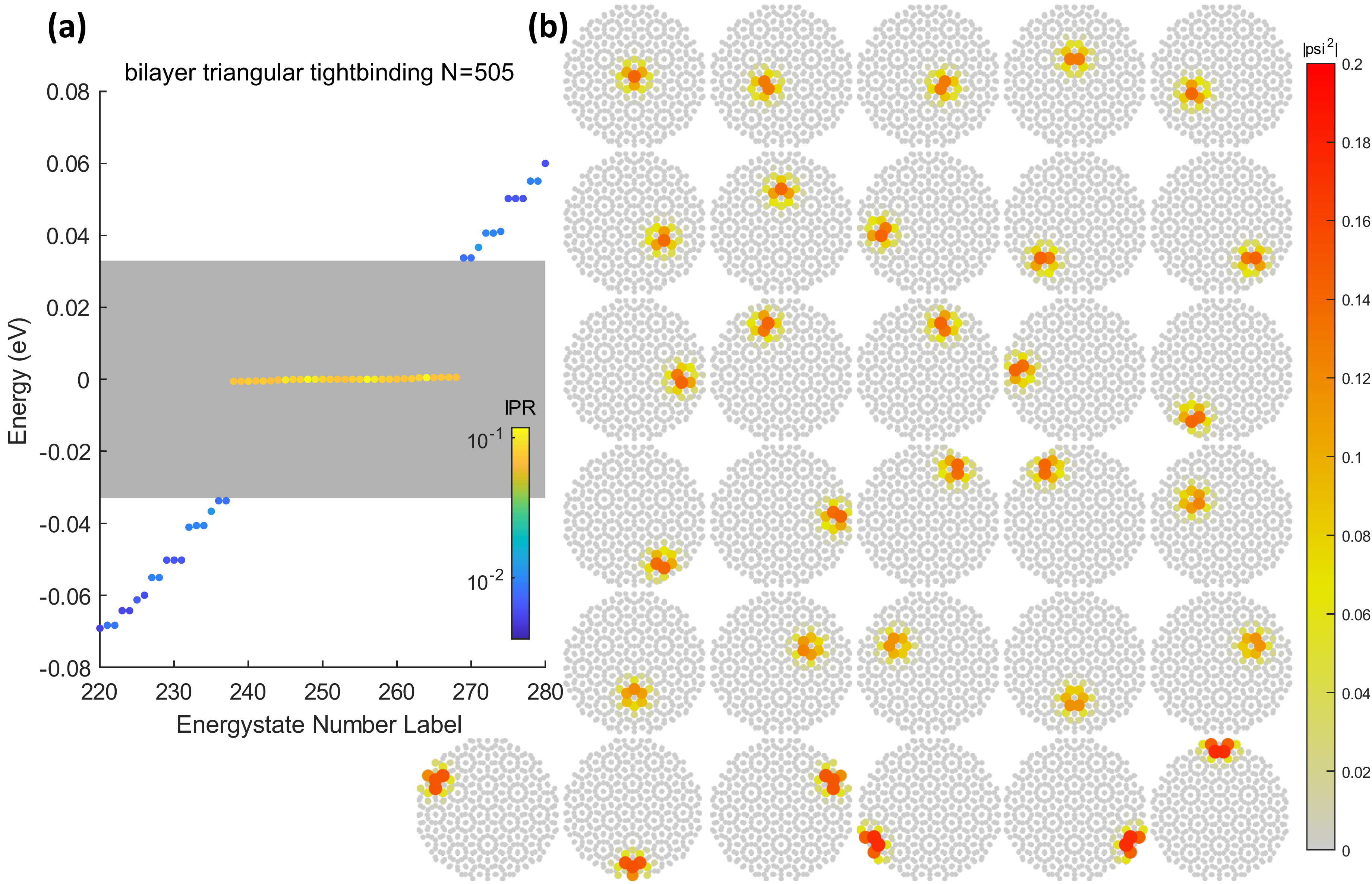}
\caption{
(a)Energy eigenvalues near zero calculated from the small sized system(N$\sim$500) for bilayer triangular lattice. The IPR value of ZESs are calculated from ELZESs which are derived from the algorithm of maximal localization. (b)Localized states calculated from the zero energy states with maximal localization algorithm.
}
\label{fig.31LS}
\end{figure*}
%%%%%%%%%%%%%%%%%%%%%%%%%%%%%%%%%%%%%%%%%%%%%%%%%%%%%%
%FIGURE%
%%%%%%%%%%%%%%%%%%%%%%%%%%%%%%%%%%%%%%%%%%%%%%%%%%%%%%

\begin{comment}
%%%%%%%%%%%%%%%%%%%%%%%%%%%%%%%%%%%%%%%%   FIGURE   %%%%%%%%%%%%%%%%%%%%%%%%%%%%%%%%%%%%%%%%%%%%%%%%%%%%%%
\begin{figure*}[t!]
\includegraphics[width=1.0\textwidth]{fig10.pdf}
\caption{
Histogram of the sum probabilities by number of considering sites for (a)bilayer traingular lattice and (i) bilayer honeycomb lattice from high symmetric localized wave function. Geometric condition for where the maximal amplitude site is described in (b) and (j), triangular and honeycomb respectively. Localized zero energy states with high symmetry(C3, C2, C6 point-group symmetry from left to right) are described in (c)-(h) for triangular lattice and (k)-(p) for honeycomb lattice. In (f)-(h) and (n)-(p), dominant sites are highlighted.
}
\label{basisnum}
\end{figure*}
%%%%%%%%%%%%%%%%%%%%%%%%%%%%%%%%%%%%%%%%   FIGURE   %%%%%%%%%%%%%%%%%%%%%%%%%%%%%%%%%%%%%%%%%%%%%%%%%%%%%%
\end{comment}

\subsection{Algorithm of maximal localization}

In this section, we show that the zero energy states, obtained by the exact diagnalization of the lattice models and the quasiband models,
give localized eigenstates in real space after suitable superposition of the nearly degenerate states. 

We consider an orthonormal set of zero energy states, $\ket{V_1}\dots \ket{V_n}$, and define a projection operator $\mathbf{P}$ spanned by them. 
Then a localized eigenstate can be constructed as

\begin{equation}
\label{LS}
\ket{LS}=\sum^n_{i=1}c_i\ket{V_i},
\end{equation}
where $\sum^n_{i=1}|c_i|^2=1$.

Next we define the spread function $\Omega$ \cite{maximallylocalizedwannier} which gives the expectation value of the spatial extent of the wave function $\ket{LS}$ as
\begin{equation}
\label{Spread function}
\Omega=<(\Delta r)^2>=\bra{LS}r^2\ket{LS}-\bra{LS}r\ket{LS}^2.
\end{equation}

In the space of $\{c_1,\cdots,c_n\}$, we find a local minimum of $\Omega$ by tracking the gradient of $\Omega$ by changing $c_i$ into $c_i + \epsilon\nabla_i{\Omega}$ with an appropriate $\epsilon$. This method is widely known as the steepest descent method. 
For the initial points, we use $\mathbf{P}\ket{\mathbf{R_{1i}}}$ ($\mathbf{P}\ket{\mathbf{R_{2j}}}$) where $\ket{\mathbf{R_{1i}}}$ ($\ket{\mathbf{R_{2j}}}$) is a wave function defined on a lattice site $i$ ($j$) in the layer 1 (layer 2).

We apply the steepest descent method to the bilayer triangular lattice with the replacement vector $\mathbf{r_d}=(a_0/2,\sqrt{3}a_0/6)$ composed of 505 lattice sites in total. 
As described in Fig.~\ref{fig.31LS}, the system has 31 zero energy states separated from the bulk states by a gap of $0.03(\mathrm{eV})$. 
Using the steepest descent method, we find exactly 31 local minimums with $\Omega<3$.

By applying the same method to the bilayer square lattice and bilayer honeycomb lattice, we again find the zero energy states, which can be described by the superposition of localized states with a spread function $\Omega<1.5$ and $\Omega<2$ respectively, which implies that the zero energy wave functions are generally described as a superposition of localized eigenstates. 

Additionally, to verify that ELZES are localized while other eigenstates with finite energy are extended states, we calculate the inverse participation ratio (IPR) for each eigenstates.~\cite{IPR1,IPR2,IPR3} The IPR for the wave function $\psi$ is explicitly defined as 

\begin{equation}
\label{IPR}
\rm{IPR}(\psi) = \sum_{\textit{i}}|\psi(\textit{i})|^{4}/(\sum_{\textit{i}}|\psi(\textit{i})|^{2})^2
,
\end{equation}
where $\textit{i}$ represents the site index. In the system with total N sites, IPR of the extended state is proportional to $O(N^{-1})$ while the IPR of the localized state is proportional to $O(1)$. The IPR of ELZES and extended states with finite energy is described in Fig.~\ref{fig.31LS}(a) which shows ELZES are localized states.

\subsection{Numerical Confirmation of the Geometrical Condition for ELZES}

In the main text, we proposed a simple geometric condition: condition for the projected position of a maximal amplitude site of the ELZES (which belongs to a single layer) to the unit cell from the other layer. In the thermodynamic limit where the number of sites is infinite, the relative location of a single site with respect to the unit cell from the other layer may uniformly be positioned (see section \ref{displacement}). From this fact, one can evaluate the ratio between the number of ZESs and the total number of eigenvalues.

The local geometrical condition for the localized state considering the maximal amplitude site is verified using the algorithm of maximal localization. Prepare a small-sized system (total 505 lattice sites for TBTL) with layer 1 fixed. The unit cell of layer 1 in the middle of the sample is highlighted with blue borders as described in Fig. 2(e) for TBTL(Fig. 3(d) for TBSL, Fig. 4(d) for TBG) in the main text. Unlike layer 1, the position of layer 2 is adjustable so we can change the location of the site $A_t$($A_s$,$A_h$) with respect to the unit cell of layer 1. Moving the location of $A_t$($A_s$,$A_h$), we investigated whether the point can be the maximal amplitude point of the zero energy state.

We start with the initial wave function $\mathbf{P}\ket{\mathbf{A_t}}$ ($\mathbf{P}\ket{\mathbf{A_s}}$,$\mathbf{P}\ket{\mathbf{A_h}}$), where $\ket{\mathbf{A_t}}$($\ket{\mathbf{A_s}}$,$\ket{\mathbf{A_h}}$) is a wave function defined on a lattice site $A_t$($A_s$,$A_h$) in layer 2. 
From the algorithm of maximal localization using the steepest descent method, we constructed the localized wave function $\ket{LS}$ from the initial wave function. If the newly constructed wave function has a maximal amplitude at $A_t$($A_s$,$A_h$), we can infer that there exists a localized zero energy state where $A_t$($A_s$,$A_h$) is the site of maximal amplitude. If the maximal amplitude site of $\ket{LS}$ is altered from $A_t$($A_s$,$A_h$), the location is forbidden for the localized zero energy state. The result of a numerical calculation is shown as the available region with blue dots and forbidden region with black dots in Fig. 2(e) for TBTL, Fig. 3(d) for TBSL, and Fig. 4(d) for TBG respectively, which is consistent with the predicted area.

ELZES in different sites can be adiabatically mapped with each other. Consider the situation where layer 1 is fixed and layer 2 can be adjusted freely. From the discussion in section \ref{displacement}, the energy eigenvalues are not changed under the displacement of layer 2. Moreover, from the quasiband method, zero energy states are shown to be isolated with other states with a gap. Therefore, the localized zero energy state remains as a zero energy state even the position of the maximal amplitude site $A_t$($A_s$,$A_h$) changes the position. Moreover, if the maximal amplitude site leaves the available area, the adjacent site of $A_t$($A_s$,$A_h$) becomes the maximal amplitude site and the amplitude from the site $A_t$($A_s$,$A_h$) is no more maximum.

\section{Twisted Bilayer Graphene (TBG) quasicrystal}
\label{section.honeycomb}

\subsection{Model Hamiltonian}
We construct a lattice model for bilayer honeycomb lattice with $30\degree$ twist taking a similar approach used for bilayer triangular lattice.
Considering that a honeycomb lattice has two sublattices $A,~B$, the lattice vector of the bilayer system can be written as
\begin{align}
    \label{eq.latticehoneycomb}
    \mathbf{R}_{1\alpha}&=n\mathbf{b}_1+m\mathbf{b}_2+\mathbf{\tau}_{\alpha},\qquad \qquad \qquad \quad (\text{layer 1}) \nonumber\\
    \mathbf{R}_{2\beta}&=n'\mathbf{b'}_1+m'\mathbf{b'}_2+\mathbf{\tau'}_{\beta}+\mathbf{r_d}+\mathbf{d_z}, \quad (\text{layer 2})
\end{align}
where $\alpha,~\beta\in \{A,B\}$ are sublattice indices and $\bf{\tau}_\alpha$ indicates the relative position of the sublattice $\alpha$ in a unit cell. 
Specifically in our model, we choose $\mathbf{b}_1=(\sqrt{3}a_0,0,0)$, $\mathbf{b}_2=(\sqrt{3}a_0/2,3a_0/2,0)$, $\mathbf{\tau}_A=(0,0,0)$, $\mathbf{\tau}_B=(\sqrt{3}a_0/2,-a_0/2,0)$ where 
$a_0\approx0.142nm$ denotes the length between nearest neighbor sites.
We note that the length of the primitive lattice vector is $\sqrt{3}a_0$ for each honeycomb layer.

Similar to the case of bilayer traingular lattice, the layer 2 is stacked right above the layer 1, displaced by $\mathbf{d_z}=(0,0,d_0)$ with $d_0\approx0.335nm$, and twisted by $30\degree$ counter-clockwisely so that $\mathbf{b'}_i=R_{z,\pi/6}\mathbf{b}_i$, $\mathbf{\tau'}_{\alpha}=R_{z,\pi/6} \mathbf{\tau}_{\alpha}$.

We consider the following Hamiltonian which is similar to the Hamiltonian in Eq.~(\ref{eq.LatticeH}),
\begin{align}
\label{eq.LatticeHhoneycomb}
H=-\sum_{\mathbf{R}_{1},\mathbf{R}_{2}}\sum_{\alpha,\beta}T(\mathbf{R}_{1\alpha}-\mathbf{R}_{2\beta})\ket{\mathbf{R}_{1\alpha}}\bra{\mathbf{R}_{2\beta}} +h.c.
\end{align}
where the transfer integral $T(\mathbf{R}_{1\alpha}-\mathbf{R}_{2\beta})$ is defined in the same way as in Eq.(\ref{eq.transfer integral}).
The above Hamiltonian also has chiral symmetry related to the layer interchange, and the relevant unitary operator is given by
\begin{align}
    \label{eq.layerchangehoneycomb}
    U=\sum_{\mathbf{R_{1\alpha}}}-\ket{R_{1\alpha}}\bra{R_{1\alpha}}+\sum_{\mathbf{R_{2\beta}}}\ket{R_{2\beta}}\bra{R_{2\beta}}.
\end{align}
Therefore the energy spectrum of the Hamiltonian is also symmetric with respect to the zero energy.

The Bloch wave function of each layer is defined as
\begin{align}
\label{eq.bloch honeycomb}
  \ket{\mathbf{k}_1,\alpha}_1=\frac{1}{\sqrt{N}}\sum_{\mathbf{R}_{1\alpha}} e^{i\mathbf{k}_1\cdot\mathbf{R}_{1\alpha}}\ket{\bf{R_{1\alpha}}}, \qquad (layer 1) \nonumber\\
  \ket{\mathbf{k}_2,\beta}_2=\frac{1}{\sqrt{N}}\sum_{\mathbf{R}_{2\beta}} e^{i\mathbf{k}_2\cdot\mathbf{R}_{2\beta}}\ket{\bf{R_{2\beta}}}, \qquad (layer 2)
\end{align}
where the momentum $\mathbf{k}_1$ ($\mathbf{k}_2$) is defined in the first BZ of the layer 1(2) and N is the number of unit cells per layer.

The matrix element of the Hamiltonian under the Bloch basis is
\begin{align}
%\begin{multline}
\label{eq.generalumklapp honeycomb}
{}_1\bra{\mathbf{k}_2,\beta}H\ket{\mathbf{k}_1,\alpha}_2 % \\
=\frac{1}{\Omega}\sum_{G_1,G_2}\delta_{\mathbf{k}_1+G_1,\mathbf{k}_2+G_2}t(\mathbf{k}_1+G_1)e^{i(-G_1 \cdot \tau_\alpha+G_2 \cdot \tau'_\beta+G_2 \cdot r_d)}
%\end{multline}
\end{align}
where $\Omega$ is the area of a unit cell (Wigner-Seitz cell) from a single layer.
$G_1$ and $G_2$ are the reciprocal lattice vectors of the layer 1 and 2, respectively. 
The Fourier component $|t(\mathbf{k}_1+G_1)|$ is defined in the same way as in the case for TBTL. 
From Eq.~(\ref{eq.generalumklapp honeycomb}), one can show that the `General Umklapp Process' is also satisfied in the bilayer honeycomb case.

\subsection{Quasiband Model with EBZ as UBZ}

We construct a quasiband model for bilayer honeycomb lattice taking a similar approach used previously.
The only difference compared to TBTL quasicrystal is that there are two eigenstates per each momentum $\bf{q}$ in the BZ because of the sublattice degrees of freedom, \textit{a} and \textit{b}.
We take the union of two first BZs (UBZ) as EBZ, which is again divided into three regions $A$, $B$, and $C$ as we did in the TBTL case. 
The quasiband Hamiltonian $\bf{H_{QB}}$ defined with the Bloch states in the EBZ is constructed as follows.

First, the Hamiltonian $\bf{H_A(q)}$ for the region $A$, which is now a $4\times4$ matrix and acts on the basis $(\ket{\mathbf{q},A}_1,\ket{\mathbf{q},B}_1,\ket{\mathbf{q},A}_2,\ket{\mathbf{q},B}_2)$, is given by
\begin{align}
\label{eq.honeycomb4*4}
\mathbf{H_A(q)} =
\left( \begin{array}{cc}
0 & \mathbf{T^h(q)}^{\dagger} \\
\mathbf{T^h(q)} & 0\\
\end{array} \right),
\end{align}
where
\begin{align}
\label{eq.honeycombin}
\mathbf{T^h(q)} =
\left( \begin{array}{cc}
T(\bf{q}) & T(\bf{q}) \\
T(\bf{q}) & T(\bf{q})\\
\end{array} \right),
\end{align}
in which $T(\bf{q})=(-1/\Omega_h) t(\bf{q})$ and $\Omega_h=\sqrt{3}(\sqrt{3}a_0)^2/2=3\sqrt{3}a_0^2/2$ is the unit cell area of a honeycomb lattice with the nearest-neighbor bond length $a_0$. 
 One can find four quasibands with energy $\pm2|T(\mathbf{q})|, 0, 0$ where two zero energy states are found per four coupled Bloch states at every momentum $\bf{q}$ in the region A. 
 The wave functions for the two zero energy states are
 $(1/\sqrt{2})(\ket{\mathbf{q},A}_1-\ket{\mathbf{q},B}_1)$ and $(1/\sqrt{2})(\ket{\mathbf{q},A}_2-\ket{\mathbf{q},B}_2)$, respectively.
 
In region B and C, 6 Bloch states are coupled, which is doubled in the number compare to the TBTL case.
Therefore $\mathbf{H_B(q)}$ and $\mathbf{H_C(q)}$ become $6\times6$ matrices. 
Specifically for the region B, $\mathbf{H_B(q)}$
acting on the basis $(\ket{\mathbf{q},A}_1,\ket{\mathbf{q},B}_1,\ket{\mathbf{q},A}_2,\ket{\mathbf{q},B}_2,\ket{\mathbf{q+G_2},A}_1,\ket{\mathbf{q+G_2},B}_1)$ is given by
\begin{align}
\label{eq.honeycomb6*6}
\mathbf{H_{B}(q)} =
\left( \begin{array}{ccc}
0 & \mathbf{T^h_1(q)}^{\dagger} & 0 \\
\mathbf{T^h_1(q)} & 0 & \mathbf{T^h_2(q)}^{\dagger} \\
0 & \mathbf{T^h_2(q)} & 0 \\
\end{array} \right),
\end{align}
where
\begin{align}
\label{eq.honeycombin}
\mathbf{T^h_1(q)} =
\left( \begin{array}{cc}
T_1(\bf{q}) & T_1(\bf{q}) \\
T_1(\bf{q}) & T_1(\bf{q})\\
\end{array} \right)
,\quad
\mathbf{T^h_2(q)} =
\left( \begin{array}{cc}
T_2(\bf{q}) & \nu T_2(\bf{q}) \\
T_2(\bf{q}) & \nu T_2(\bf{q})\\
\end{array} \right),
\end{align}
in which $T_1(\mathbf{q})=(-1/\Omega_h) t(\mathbf{q})$, $T_2(\mathbf{q})=(-1/\Omega_h) t(\mathbf{q+G_2})e^{i\mathbf{G_2}\cdot\mathbf{\tau_A'}}$, and $\nu=e^{i\mathbf{G}_2\cdot\mathbf{\tau}_B'}$.

%%%%%%%%%%%%%%%%%%%%%%%%%%%%%%%%%%%%%%%%%%%%%%%%%%%%%%%%%%%%%%%%%%%%%%%%%%%%%
%%%%%%%%%%%%%%%%%%%%%%%%%%%%%%%%%%%%%%%%%%%%%%%%%%%%%%%%%%%%%%%%%%%%%%%%%%%%%
%%%%%%%%%%%%%%%%%%%%%%%%%%%%%%%%%%%%%%%%%%%%%%%%%%%%%%%%%%%%%%%%%%%%%%%%%%%%%
%%%%%%%%%%%%%%%%%%%%%%%%%%%%%%%%%%%%%%%%%%%%%%%%%%%%%%%%%%%%%%%%%%%%%%%%%%%%%

The $6\times6$ matrix Hamiltonian also has two zero energy eigenvalues and the corresponding wave functions are $(1/\sqrt{2})(\ket{\mathbf{q},A}_1-\ket{\mathbf{q},B}_1)$ and $(1/\sqrt{2})(\ket{\mathbf{q+G_2},A}_1-\ket{\mathbf{q+G_2},B}_1)$.
One can see that the zero energy wave functions are in the same form $(1/\sqrt{2})(\ket{\mathbf{q},A}_i-\ket{\mathbf{q},B}_i)$ as in the region A, i.e., the subtraction of two Bloch wave functions in same layer $i$ and same momentum $\mathbf{q}$ but with different  sublattices. 

For the bilayer honeycomb lattice, the existence of the zero energy state in $\mathbf{H_{QB}}$ is guaranteed by the chiral symmetry $U$ about the layer interchange and the emergence of the effective sublattice change symmetry $U_{AB}^i(\mathbf{q})$ defined for a single momentum $\mathbf{q}$.
Explicitly, the sublattice change operator in the layer $i$ and the momentum $\mathbf{q}$ can be written as
\begin{align}
    \label{eq.sublattice_change}
    U_{AB}^i(\mathbf{q})=\ket{\mathbf{q},A}_i{}_i\bra{\mathbf{q},B} + \ket{\mathbf{q},B}_i{}_i\bra{\mathbf{q},A}.
\end{align}

As the sublattice change operator acts on a single layer, one can easily show that $[U,U_{AB}^i(\mathbf{q})]=0$ for arbitrary layer $i$ and momentum $\mathbf{q}$. 
As $\left[U_{AB}^i(\mathbf{q})\right]^2=I$, the eigenvalues of $U_{AB}^i(\mathbf{q})$ should be either 1 or -1. 
Moreover, the sublattice change operators commute with each other regardless of the layer and the momentum indices.

For the Hamiltonian $\bf{H_A(q)}$, the quasiband Hamiltonian has two sublattice symmetries from two layers, that is, $[\mathbf{H_A(q)},U_{AB}^1(\mathbf{q})]=[\mathbf{H_A(q)},U_{AB}^2(\mathbf{q})]=0$. Hence, the quasiband Hamiltonian can be block-diagonalized into four blocks, 
specified by the eigenvalues $\alpha$ and $\beta$ of $U_{AB}^1(\mathbf{q})$ and $U_{AB}^2(\mathbf{q})$, 
respectively, as follows

\begin{equation}
\label{eq.HA_sublattice}
\mathbf{H_A(q)} = \bigoplus_{\alpha,\beta\in\{-1,1\}} \mathbf{H_A^{(\alpha,\beta)}(q)}.
\end{equation}

The ranks of the blocks of quasiband Hamiltonian $\bf{H_A^{(-1,-1)}(q)}$, $\bf{H_A^{(-1,1)}(q)}$, $\bf{H_A^{(1,-1)}(q)}$, $\bf{H_A^{(1,1)}(q)}$ are 0, 1, 1, 2 respectively. As each block has layer chiral symmetry, zero energy states are guaranteed in $\bf{H_A^{(-1,1)}(q)}$ and $\bf{H_A^{(1,-1)}(q)}$ individually, which have an odd number of eigenvalues.

For the Hamiltonian $\bf{H_B(q)}$, the quasiband Hamiltonian loses sublattice symmetry for the layer 2. However, there are two sublattice symmetries for the layer 1 at two different momenta $\mathbf{q}$ and $\mathbf{q+G_2}$, related to the commutation relation $[\mathbf{H_B(q)},U_{AB}^1(\mathbf{q})]=[\mathbf{H_B(q)},U_{AB}^1(\mathbf{q+G_2})]=0$. Hence, $\bf{H_B(q)}$ can also be block-diagonalized into four blocks, specified by the eigenvalues $\gamma$ and $\delta$ of $U_{AB}^1(\mathbf{q})$ and $U_{AB}^1(\mathbf{q+G_2})$, respectively, so that

\begin{equation}
\label{eq.HB_sublattice}
\mathbf{H_B(q)} = \bigoplus_{\gamma,\delta\in\{-1,1\}} \mathbf{H_B^{(\gamma,\delta)}(q)}.
\end{equation}

The ranks of the blocks of quasiband Hamiltonian $\bf{H_B^{(-1,-1)}(q)}$, $\bf{H_B^{(-1,1)}(q)}$, $\bf{H_B^{(1,-1)}(q)}$, $\bf{H_B^{(1,1)}(q)}$ are 0, 1, 1, 4 respectively. As each block has layer chiral symmetry, zero energy states are guaranteed in $\bf{H_B^{(-1,1)}(q)}$ and $\bf{H_B^{(1,-1)}(q)}$ individually, which have an odd number of eigenvalues.
The same argument can also be applied to region C by simply switching the role of layer 1 and layer 2, and two zero energy states can be found from each $\mathbf{H_C(q)}$.
To sum up, the area where ZESs can exist (area Z) is the union of region A and region B for layer 1 and the union of region A and region C for layer 2.

However, the energy spectra of the full lattice model and the quasiband model with UBZ as EBZ give inconsistent results, as described in the main text. 
One is from the fact that $U_{ab}^i(\mathbf{q})$ is not a physical symmetry but an emergent symmetry owing to the presence of the momentum cutoff in the EBZ.
Thus $U_{ab}^i(\mathbf{q})$ is broken and some of the predicted ZESs are gapped when the region outside the EBZ is included.
The second is due to the large coupling strength between predicted ZESs arising from the DFR, plotted in the main text.
Contrary to TBTL and TBSL quasicrystals, 
the DFR overlap along the full boundary of the EBZ in TBG quasicrystals.
We find that the coupling strength between ZESs is comparable to the energy gap so that the quasiband model
is unreliable.

\subsection{Quasiband Model including AEBZ}
\label{cutoff}

%%%%%%%%%%%%%%%%%%%%%%%%%%%%%%%%%%%%%%%%%%%%%%%%%%%%%%
%FIGURE%
%%%%%%%%%%%%%%%%%%%%%%%%%%%%%%%%%%%%%%%%%%%%%%%%%%%%%%
\begin{figure*}[t!]
\includegraphics[width=0.75\textwidth]{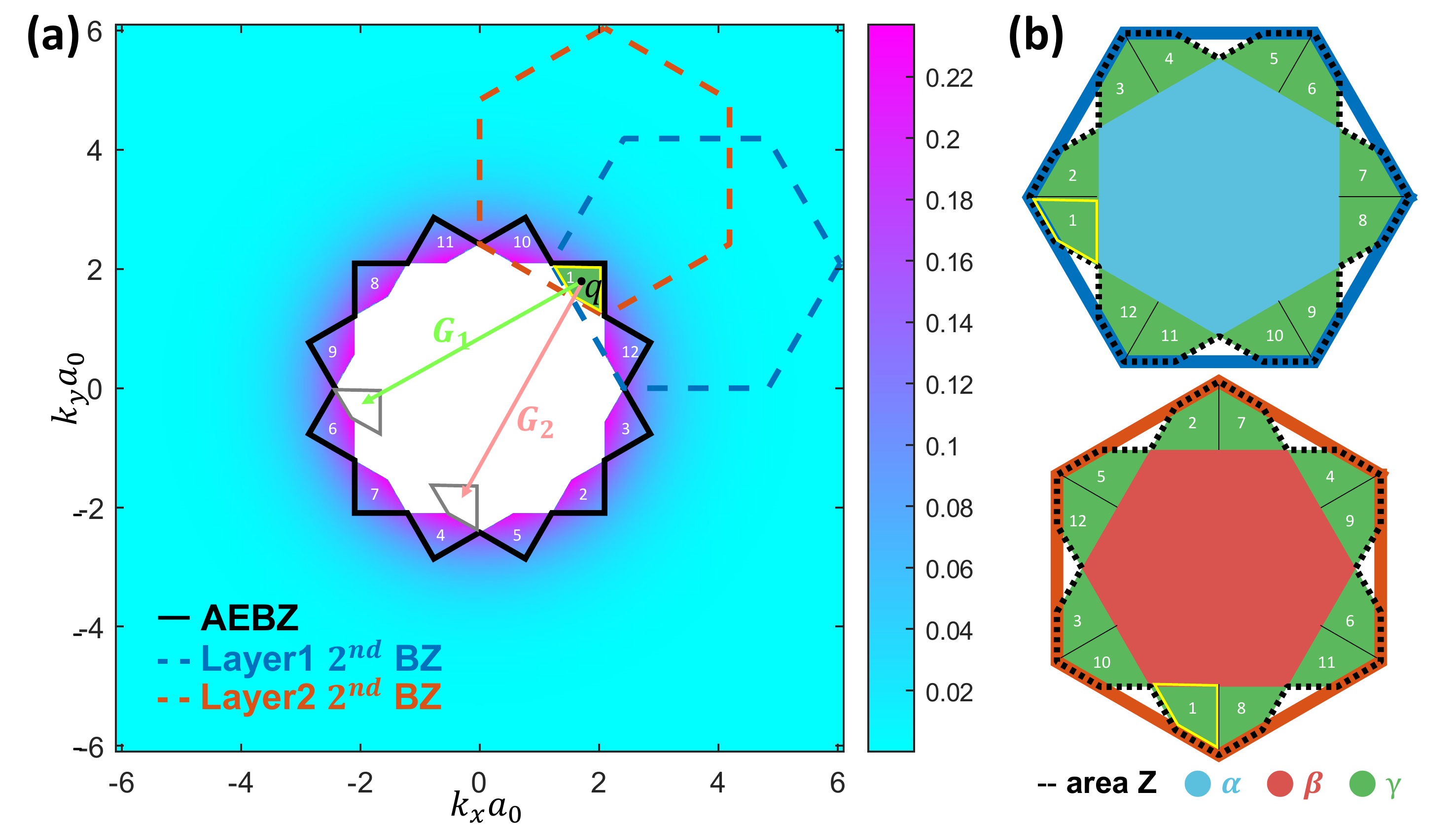}
\caption{
(a) Schematic figure describing the construction of the effective Hamiltonian with AEBZ.
The effective Hamiltonian is defined from $\bf{H-H_{QB}}$, outside of UBZ. AEBZ is divided into 12 patches. 
(b) Sector $\alpha$, $\beta$, and $\gamma$. Sector $\gamma$ is divided into 12 patches. Patches with same number but from different layers coincides inside the AEBZ. The green-filled area in (a) corresponds to the patch with number 1 in (b) for each layer which is highlighted with yellow border. Sector $\alpha$ and $\beta$ supports genuine ZESs. 
}
\label{fig.honeycombband}
\end{figure*}
%%%%%%%%%%%%%%%%%%%%%%%%%%%%%%%%%%%%%%%%%%%%%%%%%%%%%%
%FIGURE%
%%%%%%%%%%%%%%%%%%%%%%%%%%%%%%%%%%%%%%%%%%%%%%%%%%%%%%

To remedy the problem of misleading ZESs, we take into account the larger momentum region outside UBZ, referred to as the adjusted EBZ (AEBZ). Here the AEBZ is chosen so that the DFR for the new EBZ (including the UBZ and AEBZ) becomes apart from the AEBZ. Then using the extended quasiband model, we examine the influence of the AEBZ on ZESs. 

To understand the coupling between ZESs, we project $\bf{H-H_{QB}}$ into space spanned by the zero energy states from $\bf{H_{QB}}$. 
As described earlier, $\bf{H_{QB}}$ for the honeycomb lattice has zero energy states $\ket{\psi_0^i(\bf{q})}=(1/\sqrt{2})(\ket{\mathbf{q},A}_i-\ket{\mathbf{q},B}_i)$ (i=1 or 2 is the layer index) defined for the momentum $\bf{q}$ inside the area Z. If we define the projection operator to the space of the zero energy states $\ket{\psi_0^i(\bf{q})}$ as $\bf{P}$, it is sufficient to understand the effective Hamiltonian $\mathbf{H_{eff}(q)} = \bf{P(H-H_{QB})P}$. Matrix components of the effective Hamiltonian is written as

\begin{align}
%\begin{multline}
\label{eq.effectiveH1}
\bra{\psi_0^1(\mathbf{k_1})}\mathbf{H_{eff}} \ket{\psi_0^2(\mathbf{k_2})}%\\
=\frac{1}{2\Omega_h}\sum_{G_1,G_2}\delta_{\mathbf{k}_1+G_1,\mathbf{k}_2+G_2}t(\mathbf{k}_1+G_1)f(\mathbf{k}_1,G_1,\mathbf{k}_2,G_2),
%\end{multline}
\end{align}
where
\begin{align}
%\begin{multline}
\label{eq.effectiveH2}
f(\mathbf{k}_1,G_1,\mathbf{k}_2,G_2)%\\
=(e^{-iG_1 \cdot \tau_A}-e^{-iG_1 \cdot \tau_B})(e^{iG_2 \cdot \tau'_A}-e^{iG_2 \cdot \tau'_B}).
%\end{multline}
\end{align}

We note that when the momentum $\bf{q}$ is placed outside the UBZ  but still not so far from the origin so that $\bf{q}$ is placed in the second BZs of both layers, the absolute value of $f(\mathbf{k}_1,G_1,\mathbf{k}_2,G_2)$ always becomes a constant value `3' where $\mathbf{q}=\mathbf{k}_1+G_1=\mathbf{k}_2+G_2$.

Fig.~\ref{fig.honeycombband}(a) presents the absolute size of matrix component  $\mathbf{H_{eff}}$ by its mixing momentum $\mathbf{q}$. Since $\mathbf{H_{eff}}$ is projected from $\mathbf{H-H_{QB}}$, it is defined for the momentum outside the UBZ, and the area inside the UBZ is left as a blank. Similar to the Fourier component of the hopping term $|t(\mathbf{q})|$, the matrix component of $\mathbf{H_{eff}}$ also rapidly decays for the growing momentum $\mathbf{q}$. Therefore, we can apply the momentum cut-off, which we call AEBZ, and only consider relatively small momentum, as we did with EBZ previously.

We define the shape of AEBZ as follows. As UBZ is a union of two BZs of each layer, it is a star-shaped polygon with 24 edges. By extending all 24 edges outward the UBZ, we can construct a bigger star with 24 edges as described in Fig.~\ref{fig.honeycombband}(a) with black solid lines. For our calculation, we used AEBZ to be the area between the bigger star and the UBZ.

For the AEBZ defined as above, area Z can be divided into three sectors as shown in Fig.~\ref{fig.honeycombband}(b) : sector $\alpha$, $\beta$, $\gamma$. Then for all the momentum in AEBZ, they are placed in the same sector when they are mapped into the first BZ of each layer(i) by modulo $G_i$. Specifically, AEBZ is divided into 12 patches (named 1 to 12), and every patch is projected into sector $\gamma$ of both layers, to the patch with the same number as described in Fig.~\ref{fig.honeycombband}(b).

Hence, $\mathbf{H_{eff}}$ with momentum inside the AEBZ couples two ZESs from area $\gamma$ from each layer.
For the momentum $\bf{q}$ in the AEBZ, $\bf{H_{eff}}$ mixes $\ket{\psi^1_2(\mathbf{q+G_1})}$ from the layer 1 and $\ket{\psi^2_2(\mathbf{q+G_2})}$ from the layer 2 with the mixing strength $3|t(\mathbf{q})|$. Hence $\bf{H_{eff}}$ splits the zero energies from $\bf{H_{QB}}$ into two quasibands with energies $\pm3|t(\mathbf{q})|$ from the area $\gamma$.
It is worth noting that only zero energy states from the sector $\gamma$ mix together while the zero energy states from the sector $\alpha$ and $\beta$ are not affected. Therefore, DFR for extended EBZ (including AEBZ) is defined as an overlap between area $\alpha$ of layer 1 and area $\beta$ of layer 2 in the extended BZ. The new DFR is separated from the boundary of AEBZ as shown in Fig. 4(b) in the main text.

In general, there are requirements to apply our quasiband model for the larger momentum cutoff: dividing area Z into smaller sectors so each sector from both layers should be designated into the same sector in AEBZ. For all the momentum in AEBZ, the two momenta which are mapped into the BZ of each layer(i) by modulo $G_i$ should be placed in the same sectors. Moreover, the division pattern into sectors of first BZs should be identical under $30\degree$ rotation. To add with, we can disregard the case when the momentum is mapped into the region outside the area Z, since there are no ZESs outside area Z. The AEBZ proposed in Fig.~\ref{fig.honeycombband}(a) is the smallest AEBZ with AEBZ and DFR becomes apart and symmetric under $30\degree$ rotation.

We note that considering $\bf{H_{QB}}$ alone fails to capture the number of ZESs. 
However, the quasiband model taking $\bf{H_{eff}}$ into account together gives results consistent with the massive calculations of the tight-binding Hamiltonian. 
By extending the cutoff region of $\bf{H_{eff}}$, the electronic structure calculated from the quasiband model gets closer to the results from the tight-binding calculations. We verified that AEBZ successfully captures the number of ZESs by comparing with the quasiband model with larger ABEZ described in Section~\ref{largerAEBZ}. The result is described in Section~\ref{largerAEBZ} gives a consistent result with the AEBZ introduced in  Fig.~\ref{fig.honeycombband}(a). This implies that the number of ZESs states does not change as we enlarge the range of the momentum cutoff. 

The ratio of zero energy states to the total number of states can also be predicted in TBG. Since the honeycomb lattice has two sublattices, a single point in the BZ of each layer was mapped into two Bloch states and the total number of states are proportional to the area of two first BZs multiplied by two. However there are only one zero energy Bloch state $\ket{\psi_0^i(\bf{q})}=(1/\sqrt{2})(\ket{\mathbf{q},A}_i-\ket{\mathbf{q},B}_i)$ per single quasi momentum in each layer which is located in area Z. Also  considering the effective Hamiltonian, actual zero energy states are originated from the zero energy Bloch states in sectors $\alpha$ and $\beta$. Hence the number of total zero energy states is proportional to the sum of area $\alpha$ and $\beta$. In conclusion, the ratio of zero energy states in TBG $P_{TBG}=2-\sqrt{3}\simeq0.268$ is predicted.

\subsection{Quasiband Model considering larger AEBZ}
\label{largerAEBZ}

In this section, we consider larger AEBZ and give validation for the number of ZESs calculated from the quasiband model with AEBZ. In the case of TBSL, the fact that region A (which decides the size of a gap) and DFR is attached was not an issue since the strength of mixing between ZESs was always smaller than the size of a gap. However, in the case of TBG, the size of the gap is decided from the coupling between inaccurate ZESs from sector $\gamma$ with the momentum placed in AEBZ. Therefore, unlikely as TBSL, the size of the coupling between ZESs from sector $\alpha$ ad $\beta$ is equivalent to the size of a gap at the point where DFR and AEBZ attach as in Fig. 4(b) in the main text. Nevertheless, we prove that the coupling between ZESs is negligible and ZESs are isolated in the gap even the coupling for larger momentum outside the AEBZ is perturbed. 

\begin{figure*}[t!]
\includegraphics[width=1\textwidth]{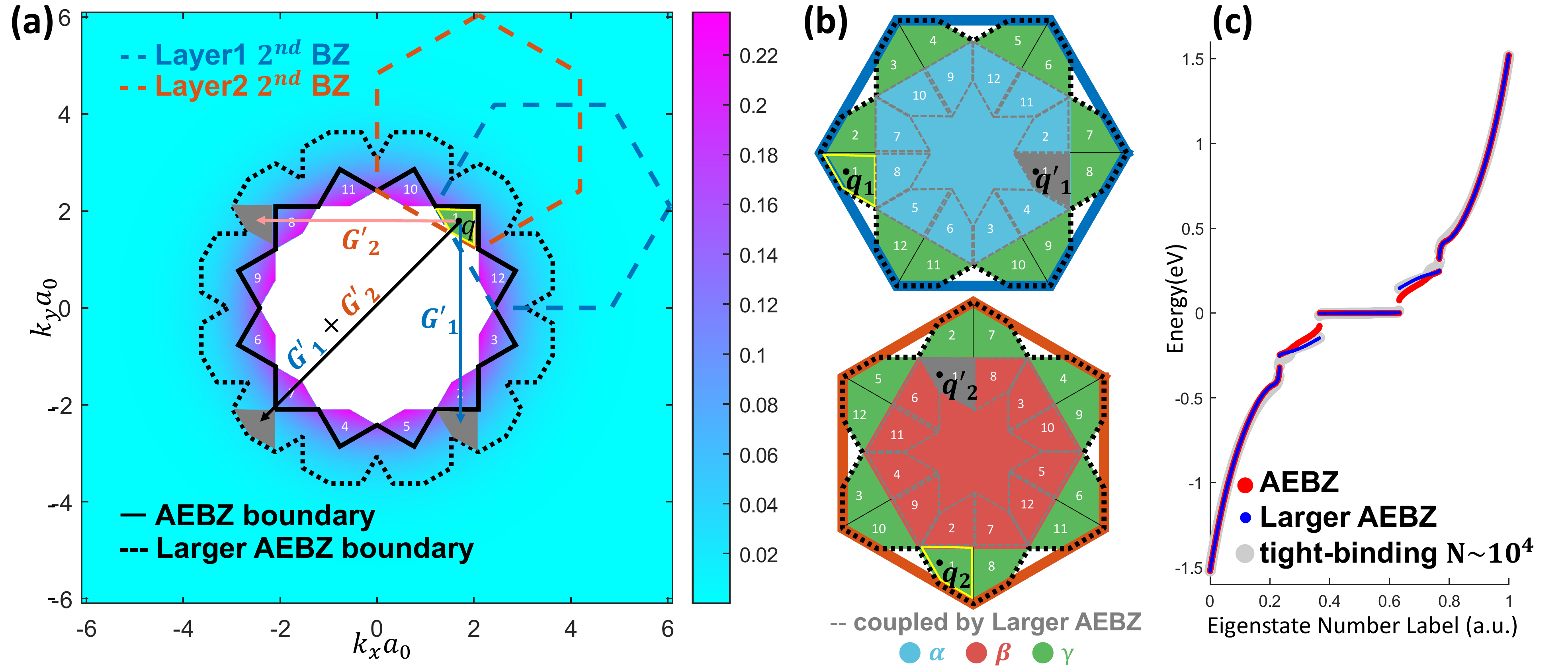}
\caption{
(a) Description for larger AEBZ. Due to the appearance of larger ABEZ, 
(b) Momentum which are in consideration due to the larger AEBZ. Not only the patch 1 in sector $\gamma$ with yellow border but also the area with grey filling in sector $\alpha$ and $\beta$ should be considered.
(c) Energy spectra calculated from a finite-size tight-binding Hamiltonian (grey), quasiband model for AEBZ (red), and quasiband model for larger AEBZ (blue).
}
\label{fig.largerAEBZ}
\end{figure*}

We consider a larger AEBZ boundary as illustrated in Fig.~\ref{fig.largerAEBZ}(a). As the AEBZ is composed of 12 identical patches with a vertex with the right angle, we rotate the patches by 90\degree, 180\degree, and 270\degree and connect so their right angle touch with each other and fill the plane tightly. Therefore, there are 48 patches, and we define the total area as the larger AEBZ.

The larger AEBZ couples two ZESs from section $\gamma$, one ZES from section $\alpha$ and one ZES from section $\beta$. Specifically, let us investigate the coupling between ZESs by considering the patch with the number 1 as described in Fig.~\ref{fig.largerAEBZ}(a) and Fig.~\ref{fig.largerAEBZ}(b). Let us define $\mathbf{q_1}$ and $\mathbf{q_2}$ identical to $\mathbf{q}$ by modulo $G_1$ and $G_2$, $\mathbf{q_1'}$ identical to $\mathbf{q}+G_2'$ by modulo $G_1$, and $\mathbf{q_2'}$ identical to $\mathbf{q}+G_1'$ by modulo $G_2$, while $G_1'$ and $G_2'$ are reciprocal lattice vectors of layer 1 and layer 2 as described in  Fig.~\ref{fig.largerAEBZ}(a). Then $\mathbf{q_1}$ and $\mathbf{q_2}$ couples with momentum $\mathbf{q}$, $\mathbf{q_1'}$ and $\mathbf{q_2}$ couples with momentum $\mathbf{q}+G_2'$, $\mathbf{q_1}$ and $\mathbf{q_2'}$ couples with momentum $\mathbf{q}+G_1'$, and $\mathbf{q_1'}$ and $\mathbf{q_2'}$ couples with momentum $\mathbf{q}+G_1'+G_2'$. Therefore, from equation \ref{eq.effectiveH1} and \ref{eq.effectiveH2}, $\mathbf{H_{eff}}$ which acts on ($\ket{\mathbf{q_1}}$, $\ket{\mathbf{q_1'}}$, $\ket{\mathbf{q_2}}$, $\ket{\mathbf{q_2'}}$) is written as

\begin{align}
\label{eq.largerAEBZ}
\mathbf{H_{eff}(q)} =
\left( \begin{array}{cc}
0 & \mathbf{T(q)} \\
\mathbf{T(q)}^{\dagger} & 0
\end{array} \right),
\end{align}

where
\begin{align}
\label{eq.T}
\mathbf{T(q)} =
\left( \begin{array}{cc}
(1-w)(1-w^2)t(\mathbf{q}) & (1-w^2)(1-w^2)t(\mathbf{q}+G_1') \\
(1-w)(1-w)t(\mathbf{q}+G_2') & (1-w^2)(1-w)t(\mathbf{q}+G_1'+G_2')\\
\end{array} \right)
\end{align}

and $w=e^{i{2\pi \over 3}}$. The value of $f(\mathbf{k}_1,G_1,\mathbf{k}_2,G_2)$ from (1,1) component and (2,2) component of $\mathbf{T(q)}$ is always complex conjugate regardless of the number of the patch. It also holds for (1,2) and (2,1). 

To note with, the size of $t(\mathbf{q})$ is determined by the size of momentum $\mathbf{q}$, so the absolute value of matrix components of $\mathbf{T(q)}$ are similar in size. Moreover, as the hopping term $T(\mathbf{R})$ is a real value, $t(\mathbf{-q})^*=t(\mathbf{q})$ holds as $t(\mathbf{q})$ is a Fourier component of $T(\mathbf{R})$. Therefore, $\mathbf{T(q)}$ can be written as

\begin{align}
\label{eq.approx}
\mathbf{T(q)} = \mathbf{T_0} + \mathbf{\Delta T} =
\left( \begin{array}{cc}
t_1 & t_2 \\
t_1^* & t_2^* \\
\end{array} \right)
+\mathbf{\Delta T}
\end{align}

where $|t_1|=|t_2|$ and the size of matrix components of $\mathbf{\Delta T}$ much smaller than $|t_1|$. $\mathbf{H_{eff}(q)}$ has four eigenvalues, where two of them are the square root of the eigenvalues of  $\mathbf{T(q)T(q)^{\dagger}}$ and the other two are also the square root of the eigenvalues of $\mathbf{T(q)^{\dagger}T(q)}$. We can find the eigenvalues of  $\mathbf{H_{eff}(q)}$ with the perturbation theory as treating $\mathbf{\Delta T}$ as a perturbation. 
If we expand $\mathbf{T(q)^{\dagger}T(q)}$ using equation ~\ref{eq.approx}, we can find

\begin{align}
\label{eq.perturb}
\mathbf{T(q)T(q)^{\dagger}} = (\mathbf{T_0 + \Delta T})(\mathbf{T_0 + \Delta T})^{\dagger} = \mathbf{T_0 T_0^{\dagger}} + \mathbf{T_0 \Delta T^{\dagger}} + \mathbf{\Delta T T_0^{\dagger}} + O(\mathbf{\Delta T}^2), \nonumber\\
\mathbf{T(q)^{\dagger}T(q)} = (\mathbf{T_0 + \Delta T})^{\dagger}(\mathbf{T_0 + \Delta T}) = \mathbf{T_0^{\dagger} T_0} + \mathbf{T_0^{\dagger} \Delta T} + \mathbf{\Delta T^{\dagger} T_0} + O(\mathbf{\Delta T}^2).
\end{align}

The zeroth order of eigenvalues are the square root of the eigenvalues of $\mathbf{T_0 T_0^{\dagger}}$ and $\mathbf{T_0^{\dagger} T_0}$. As both matrices have eigenvalue 0 and $2(|t_1|^2+|t_2|^2)$, $\mathbf{H_{eff}(q)}$ has two zero energy state in the zeroth order of $\mathbf{\Delta T}$. Let us define the eigenvector of  $\mathbf{T_0 T_0^{\dagger}}$ and $\mathbf{T_0^{\dagger} T_0}$ with zero eigenvalue as $v_1$ and $v_2$. %
To find the first order of the zero energy states, we compute $v_1^{\dagger}(\mathbf{T_0 \Delta T^{\dagger}} + \mathbf{\Delta T T_0^{\dagger}})v_1$ and $v_2^{\dagger}(\mathbf{T_0^{\dagger} \Delta T} + \mathbf{\Delta T^{\dagger} T_0})v_2$. However, as $\mathbf{T_0^{\dagger}}v_1 = 0$ and $\mathbf{T_0}v_2 = 0$, the first order also becomes zero. This implies that zero energy is protected over $O(\mathbf{\Delta T}^2)$ even the larger AEBZ is introduced and coupled the ZESs from section $\alpha$ and $\beta$.

In conclusion, the coupling between ZESs is only valid for the momentum outside the larger AEBZ, which is separated far away from the AEBZ boundary which determines the size of a gap, and the energy splitting of ZESs are negligible. This fact is verified by comparing the result of quasiband model with AEBZ, larger AEBZ, and the result of direct diagonalization of the tight-binding model as shown in Fig.~\ref{fig.largerAEBZ}(c).

%%%%%%%%%%%%%%%%%%%%%%%%%%%%%%%%%%%%%%%%%%%%%%%%%%%%%%
%FIGURE%
%%%%%%%%%%%%%%%%%%%%%%%%%%%%%%%%%%%%%%%%%%%%%%%%%%%%%%

\section{Independence of the spectrum on $\mathbf{r_d}$}
\label{displacement}
For the infinitely large system, we state that the system is independent of the displacement vector $r_d$. In other words, a system with displacement $r_d$ is identical to the system with no displacement.

The statement can be proven by the following observation,\\
$\forall \epsilon>0$, $\exists \,n,m,n',m'\in \mathbf{Z}$ s.t. $|r_d + \tilde{R}_{1\alpha}-\tilde{R}_{2\beta}|<\epsilon$, where $\tilde{R}_{1\alpha}=nb_1+mb_2, \tilde{R}_{2\beta}=n'b'_1+m'b'_2$

For simplicity, set $a_0=1$ for triangular lattice and $\sqrt{3}a_0=1$ for honeycomb lattice, distance can be written in the form
\begin{align}
\nonumber
r_d + \tilde{R}_{1\alpha}-\tilde{R}_{2\beta}=r_d+\begin{pmatrix}1&  \frac{1}{2} \\0 & \frac{\sqrt{3}}{2} \end{pmatrix}\begin{pmatrix}n \\m \end{pmatrix}-\begin{pmatrix}\frac{\sqrt{3}}{2} &0 \\\frac{1}{2} & 1\end{pmatrix}\begin{pmatrix}n' \\m' \end{pmatrix}
\end{align}

Therefore it is sufficient to find the integer sets $n, m, n', m'$ which satisfies

\begin{align}
\arrowvert r_d' +\frac{1}{\sqrt{3}}\begin{pmatrix}2&1 \\-1&1 \end{pmatrix}\begin{pmatrix}n \\m \end{pmatrix}-\begin{pmatrix} n'\\ m' \end{pmatrix} \arrowvert<\epsilon'
\nonumber
\end{align}
with 
\begin{align}
\nonumber
r_d'=\frac{1}{\sqrt{3}}\begin{pmatrix}2&0 \\-1&\sqrt{3} \end{pmatrix}r_d.
\end{align}
From LU decomposition, we can treat the 2-D aperiodic problem similar to 1-D problem.
\begin{align}
\arrowvert r_d' +\frac{1}{\sqrt{3}}\begin{pmatrix}1&0 \\-2&1 \end{pmatrix}\begin{pmatrix}-n+m \\3m \end{pmatrix} -\begin{pmatrix}n'\\m'\end{pmatrix}\arrowvert<\epsilon'
\nonumber
\end{align}
Existence of $n,m,n',m'$ is trivial since the integer multiple of a irrational number is dense in [0,1].

For the square lattice, the statement is trivial since $\sqrt{2}n$ (mod 1) is dense in [0,1] for integer $n$. From this observation, if the two centers of each layer do not commensurate by the displacement vector $r_d$, there exist two sites from each layer infinitesimally close to each other in the x-y plane. Therefore the system with displacement can be considered as a translation of a system with no displacement.

\section{Determining Edge Modes from Finite Size Effect}
\label{edge}

%%%%%%%%%%%%%%%%%%%%%%%%%%%%%%%%%%%%%%%%%%%%%%%%%%%%%%
%FIGURE%
%%%%%%%%%%%%%%%%%%%%%%%%%%%%%%%%%%%%%%%%%%%%%%%%%%%%%%
\begin{figure}[t!]
\includegraphics[width=8.5cm]{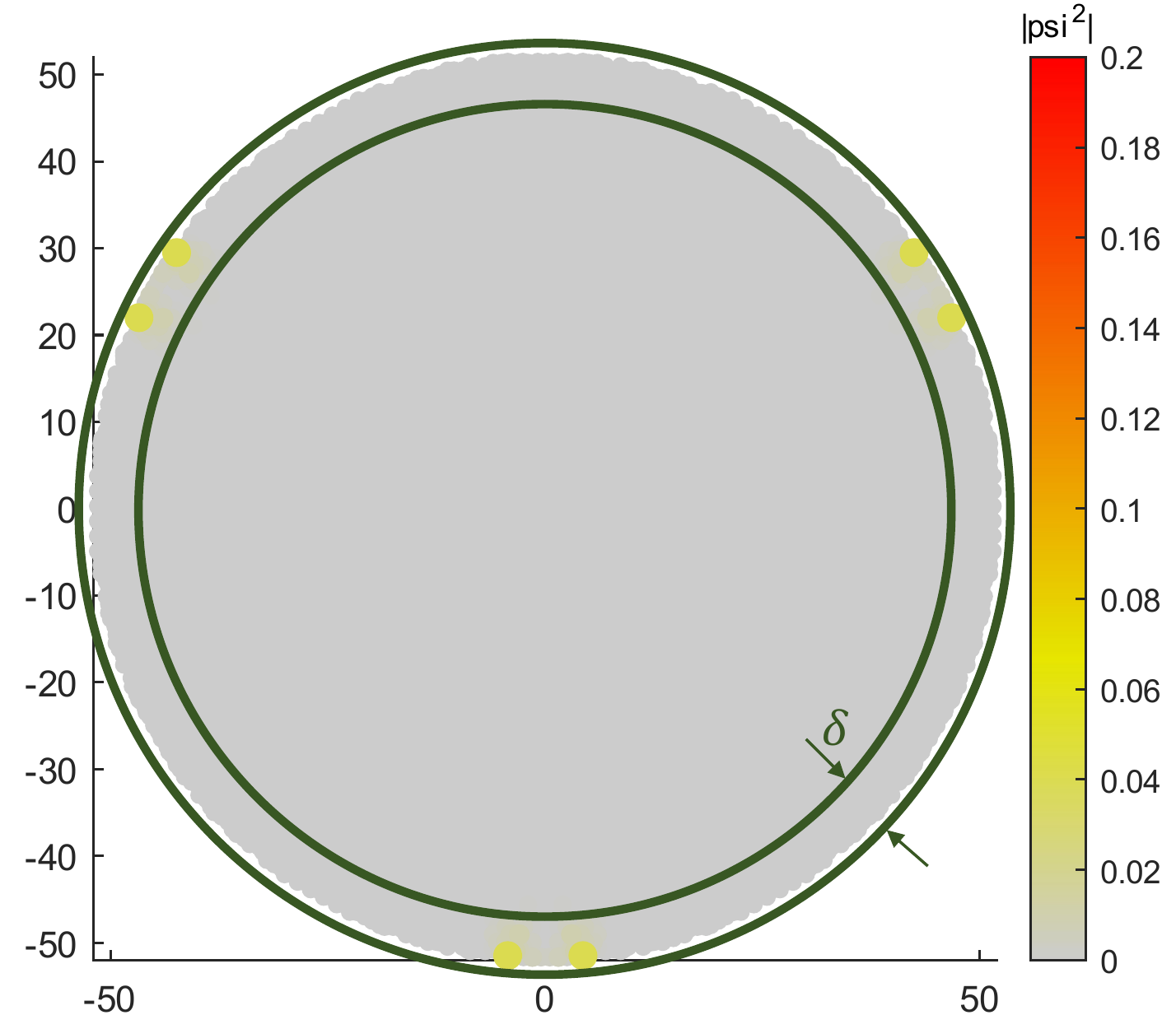}
\caption{
 Identifying the edge state. The figure above is an edge state from the bilayer triangular lattice with $r_{\star}/r_0=1$. 
}
\label{fig.edgecheck}
\end{figure}
%%%%%%%%%%%%%%%%%%%%%%%%%%%%%%%%%%%%%%%%%%%%%%%%%%%%%%
%FIGURE%
%%%%%%%%%%%%%%%%%%%%%%%%%%%%%%%%%%%%%%%%%%%%%%%%%%%%%%
Even dealing with large systems with sites $\approx10^4$ in the massive tight-binding calculation, edge effect occurs as a finite size effect.

Such edge states are distinguished with the bulk state by calculating the value of IPR since the IPR of a bulk state scales as $O(N^{-2})$ while the IPR of an edge state scales as $O(N^{-1})$. Moreover, through the tight-binding calculation, ELZES may be found near the edge with energy deviated from zero. Such states are located in the sites which satisfy the geometrical conditions for the localized states but are located near the boundary. IPR of these states is scaled as $O(1)$. However, the ratio of the number of edge states over a total number of states scales as $O(N^{-1/2})$ which converges to zero for sufficiently large N.

To check if the given state is an edge state, we define the edge region as a ring with width $\delta$ along the boundary as shown in Fig. \ref{fig.edgecheck}. If the probability inside the edge region is bigger than $P_c$, we identified such a state as an edge state. In our research, we used $\delta=5a_0$, $P_c=0.9$.

%\bibliographystyle{unsrt} 
%\bibliography{main}